\documentclass[article]{jss}

\usepackage[english]{babel}
\usepackage[utf8x]{inputenc}
\usepackage[T1]{fontenc}

\usepackage{thumbpdf,lmodern}

\usepackage{framed}

\usepackage[T1]{fontenc}
\usepackage{amsmath}

\DeclareMathOperator{\BE}{Beta}

\graphicspath{{fig/}}

\author{Jure Demšar\\University of Ljubljana
    \And Grega Repovš\\University of Ljubljana
    \And Erik Štrumbelj\\University of Ljubljana}
\Plainauthor{Jure Demšar, Grega Repovš, Erik Štrumbelj}

\title{\pkg{bayes4psy} -- an Open Source \proglang{R} Package for Bayesian Statistics in Psychology}
\Plaintitle{bayes4psy - an Open Source R Package for Bayesian Statistics in Psychology}
\Shorttitle{\pkg{bayes4psy}}

\Abstract{
    Research in psychology generates interesting data sets and unique statistical modelling tasks. However, these tasks, while important, are often very specific, so appropriate statistical models and methods cannot be found in accessible Bayesian tools. As a result, the use of Bayesian methods is limited to those that have the technical and statistical fundamentals that are required for probabilistic programming. Such knowledge is not part of the typical psychology curriculum and is a difficult obstacle for psychology students and researchers to overcome. The goal of the \pkg{bayes4psy} package is to bridge this gap and offer a collection of models and methods to be used for data analysis that arises from psychology experiments and as a teaching tool for Bayesian statistics in psychology. The package contains Bayesian t-test and bootstrapping and models for analyzing reaction times, success rates, and colors. It also provides all the diagnostic, analytic and visualization tools for the modern Bayesian data analysis workflow.
}

\Plainkeywords{R, Bayesian statistics, psychology, reaction time, success rate, Bayesian t-test, color analysis, linear model, bootstrap}

\Address{
  Jure Demšar\\
  Faculty of Computer and Information Science\\
  University of Ljubljana\\
  Večna pot 113\\
  1000 Ljubljana, Slovenia\\
  E-mail: \email{jure.demsar@fri.uni-lj.si}
}

\begin{document}

\section[Introduction]{Introduction}
\label{sec:intro}

Through the development of specialized probabilistic models Bayesian data analysis offers a highly flexible, intuitive and transparent alternative to classical statistics. Bayesian approaches were on the sidelines of data analysis throughout much of the modern era of science. Mostly due to the fact that computations required for Bayesian analysis are usually quite complex. But computations that were only a decade or two ago too complex for specialized computers can nowadays be executed on average desktop computers. In part also due to modern Markov chain Monte Carlo (MCMC) methods that make computations tractable for virtually all parametric models. This, along with specialized probabilistic programming languages for Bayesian modelling -- such as \proglang{Stan} \citep{Carpenter2017Stan} and \proglang{JAGS} \citep{Plummer2003Jags} -- drastically increased the accessibility and usefulness of Bayesian methodology for data analysis. Indeed, Bayesian data analysis is steadily gaining momentum in the 21\textsuperscript{st} century \citep{Gelman2014Bayesian,Kruschke2014BayesianDataAnalysis,McElreath2018Rethinking}, mainly so in natural and technical sciences. Unfortunately, the use of Bayesian data analysis in social sciences remains scarce, most likely due to the steep statistical and technical learning curve of Bayesian methods.

There are many advantages of Bayesian data analysis, such as its ability to work with missing data and combine prior information with data in a natural and principled way. Furthermore, Bayesian methods offer high flexibility through hierarchical modelling and calculated posterior distribution, while calculated posterior parameter values can be used as easily interpretable alternatives to p-values -- Bayesian methods provide very intuitive answers, such as ``the parameter $\mu$ has a probability of 0.95 of falling inside the [-2, 2] interval'' \citep{Dunson2001Commentary,Gelman2014Bayesian,Kruschke2014BayesianDataAnalysis,McElreath2018Rethinking}. One of the social sciences that could benefit the most from Bayesian methodology is psychology. The majority of the data that arises in psychological experiments, such as reaction times, success rates, and colors, can be analyzed in a Bayesian manner by using a small set of probabilistic models. Bayesian methodology could also alleviate the replication crisis that is pestering the field of psychology \citep{OpenScience2015Estimating,Schooler2014Metascience,Stanley2018Meta}.

The ability to replicate scientific findings is of paramount importance to scientific progress \citep{Baker2016Crisis,McNutt2014Reproducibility,Munafo2017Manifesto}. Unfortunately, more and more replication attempts report that they had failed to reproduce original results and conclusions \citep{Amrhein2019Scientists,OpenScience2015Estimating,Schooler2014Metascience}. This so-called replication crisis is harmful not only to the authors but to science itself. A recent attempt to replicate 100 studies from three prominent psychology journals \citep{OpenScience2015Estimating} showed that only approximately a third of studies that claimed statistical significance (p value lower than 0.05) also showed statistical significance in replication. Another recent study \citep{Camerer2018Evaluating} tried to replicate systematically selected studies in the social sciences published in Nature and Science between 2010 and 2015, replication attempts were successful only in 13 cases out of 21.

The main reasons behind the replication crisis seem to be poor quality control in journals, unclear writing and inadequate statistical analysis \citep{Hurlbert2019Coup,Wasserstein2016ASA,Wasserstein2019Moving}. One of the main issues lies in the desire to claim statistical significance through p-values. Many manuscripts published today repeat the same mistakes even though prominent statisticians prepared extensive guidelines about what to do and mainly what not to do  \citep{Hubbard2015Corrupt,Wasserstein2016ASA,Wasserstein2019Moving,Ziliak2019GVaules}. Reluctance to adhere to modern statistical practices has led scientist to believe that a more drastic shift in statistical thinking is needed, and some believe that it might come in the form of Bayesian statistics \citep{Dunson2001Commentary,Gelman2014Bayesian,Kruschke2014BayesianDataAnalysis,McElreath2018Rethinking}.

The \pkg{bayes4psy} \proglang{R} package provides a state-of-the art framework for Bayesian analysis of psychological data. It incorporates a set of probabilistic models that can be facilitated for analysis of data that arises during many types of psychological experiments. All models are pre-compiled, meaning that users do not need any specialized software or skills (e.g. knowledge of probabilistic programming languages), the only requirements are the \proglang{R} programming language and very basic programming skills (same skills as needed for classical statistical analysis in \proglang{R}). Besides the probabilistic models, the package also incorporates the diagnostic, analytic and visualization tools required for modern Bayesian data analysis. As such the \pkg{bayes4psy} package represents a bridge into the exciting world of Bayesian statistics for all students and researches in the field of Psychology.

\section{Models and methods}
\label{sec:models}

For statistical computation, that is, sampling from the posterior distributions, the \pkg{bayes4psy} package utilizes \proglang{Stan} \citep{Carpenter2017Stan}. \proglang{Stan} is a state-of-the-art platform for statistical modeling and high-performance statistical computation and offers full Bayesian statistical inference with MCMC sampling. It also offers user friendly interfaces with most programming languages used for statistical analysis, including \proglang{R}. \proglang{R} \citep{R2017Language} is one of the most powerful and widespread programming languages for statistics and visualization. Visualizations in the \pkg{bayes4psy} package are based on the \pkg{ggplot2} package \citep{Wickham2009ggplot}.

\subsection{Priors}

In Bayesian statistics we use prior probability distributions (priors) to express our beliefs about the parameters before any evidence (data) is taken into account. Priors represent an elegant way of intertwining previous knowledge with new facts about the domain of analysis. Prior distributions are usually based on previously conducted and verified research or on knowledge provided by the domain experts. If such data is not available, we usually resort to our own weakly informative, vague prior knowledge.

In the \pkg{bayes4psy} package users are able to express prior knowledge by putting prior distributions on all of the model's parameters. Users can express their knowledge by using uniform, normal, gamma, or beta distributions. If users do not specify any prior knowledge about the model's parameters, then flat/improper priors are put on those parameters. For details see the practical illustrations of using the \pkg{bayes4psy} package in Section \ref{sec:illustrations}.

\subsection{Bayesian t-test}
\label{sec:ttest}

\begin{figure}[ht]
    \centering
    \includegraphics[width=0.75\textwidth]{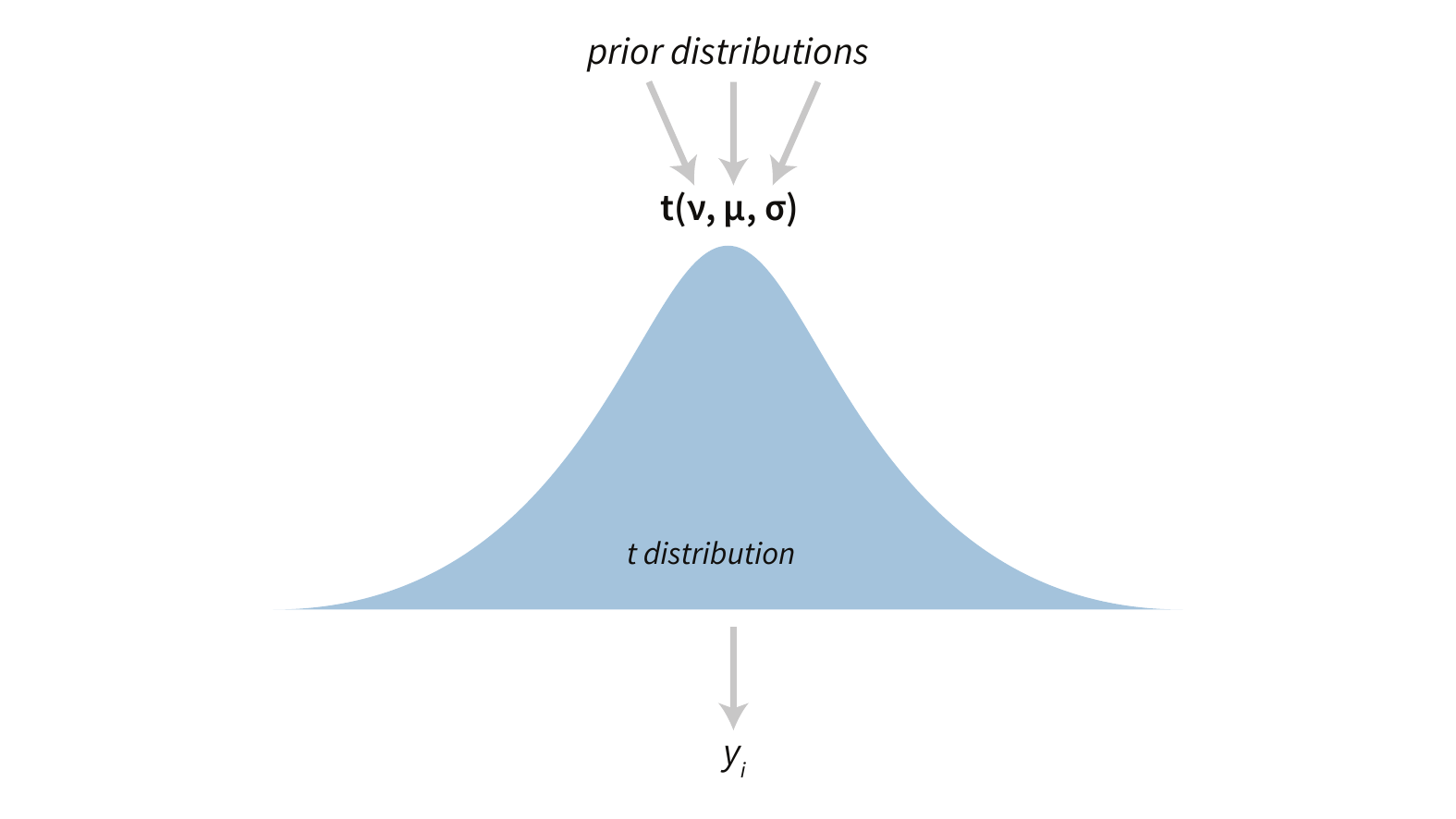}
    \caption{{\bf Visualization of the Bayesian t-test.} The model has three parameters -- degrees of freedom $\nu$, mean $\mu$ and variance $\sigma$. $y_i$ denotes $i$-th datum in the provided data set.}
    \label{fig:ttest}
\end{figure}

The t-test is probably the most commonly used hypothesis test. We added the Bayesian version of t-test to the \pkg{bayes4psy} package. The t-test is based on Kruschke's model \citep{Kruschke2013BayesianSupersedesTTest,Kruschke2014BayesianDataAnalysis}. The Bayesian t-test uses a scaled and shifted Student's t distribution (\figurename~\ref{fig:ttest}). This distribution has three parameters -- degrees of freedom ($\nu$), mean ($\mu$) and variance ($\sigma$)).

There are some minor differences between our implementation and Kruschke's. Instead of pre-defined vague priors for all parameters, we can define custom priors for the $\nu$, $\mu$ and $\sigma$ parameters. Since Kruschke's main goal was the comparsion between two groups, his implementation models two data sets simultaneously. Our implementation is more flexible, users can model several data sets individually and then make pairwise comparisons or a simultaneous cross comparison between multiple fits. We illustrate the use of the t-test in Section \ref{sec:stroop}.

\subsection{Model for analyzing reaction times}

Psychological experiments typically have a hierarchical structure -- each subject performs the same test for a number of times, several subjects are then grouped together on their characteristics (e.g. by age, sex, health) and the final statistical analysis is then conducted at the group level. Such structure is ideal for Bayesian hierarchical modelling \citep{Kruschke2014BayesianDataAnalysis}.

Our subject-level reaction time model is based on the exponentially modified normal distribution. This distribution has proven to be a suitable interpretation for the long tailed data that arise in reaction time measurements. To model the data at the group level we put hierarchical normal priors on all parameters of the subject-level exponentially modified normal distribution.

\begin{figure}[ht]
    \centering
    \includegraphics[width=\textwidth]{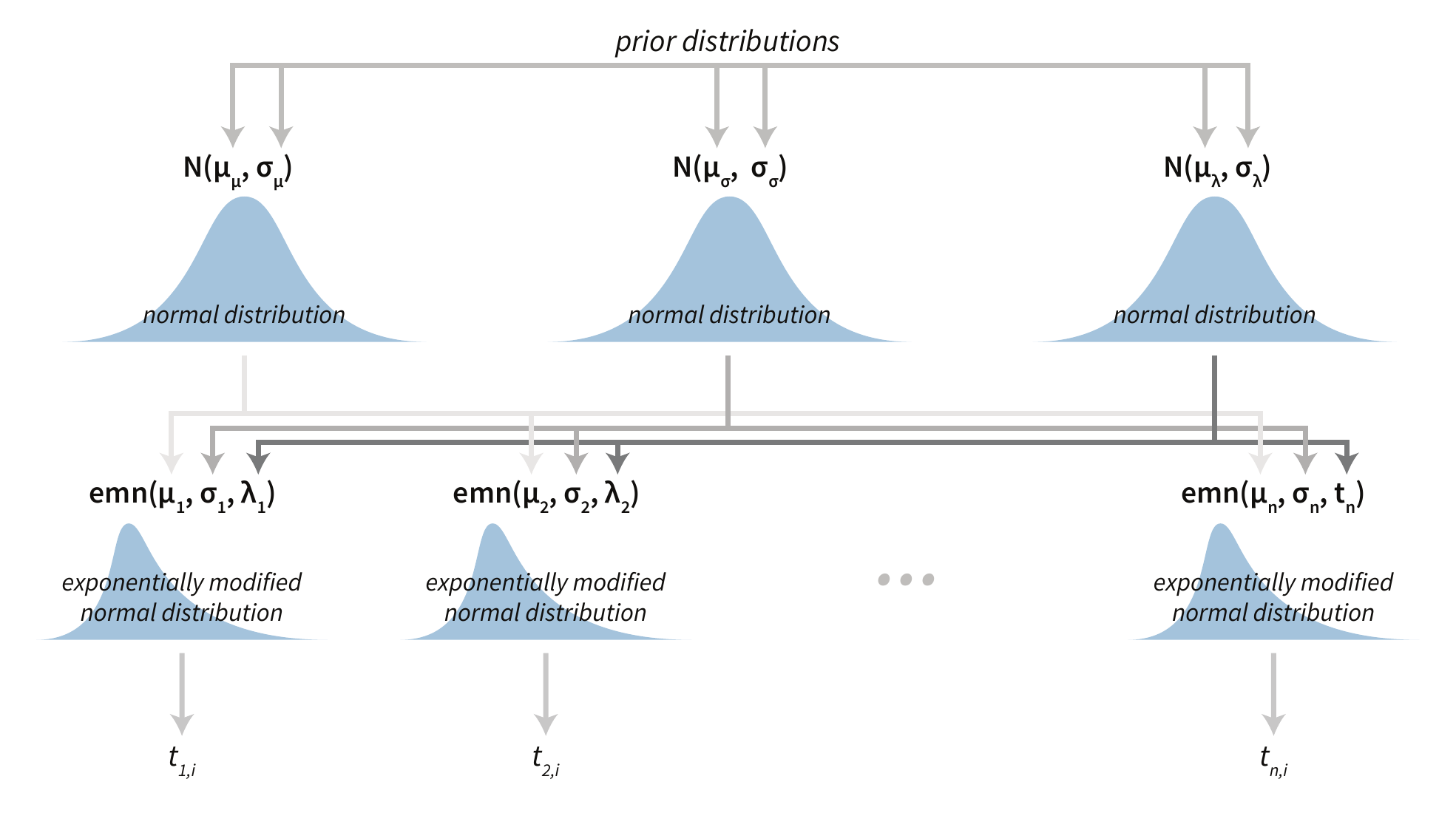}
    \caption{{\bf Visualization of the Bayesian reaction time model.} The model has a hierarchical structure, reaction times belonging to each individual subject ($t_{n,i}$ depicts $i$-th reaction time of the subject $n$) are used to construct exponentially modified normal distributions at the subject level. Parameters of subject level distributions are then connected at the group level by using normal distributions, which can then be used for group level analysis.}
    \label{fig:reaction_time}
\end{figure}

The subject level parameters are thus $\mu_i$, $\sigma_i$ and $\lambda_i$, where $i$ is the subject index. And hierarchical normal priors on these parameters are $\mathcal{N}(\mu_m, \sigma_m)$ for the $\mu$ parameter, $\mathcal{N}(\mu_s, \sigma_s)$ for the $\sigma$ parameter and $\mathcal{N}(\mu_l, \sigma_l)$ for the $\lambda$ parameter. \figurename~\ref{fig:reaction_time} is a graphical representation of the Bayesian reaction time model. For a practical application of this model see Section \ref{sec:flanker}.

\subsection{Model for analyzing success rates}

The success rate model is based on the Bernoulli-Beta model that is found in most Bayesian statistics textbooks \citep{Gelman2014Bayesian,Kruschke2014BayesianDataAnalysis,McElreath2018Rethinking}. This model is used for modelling binary data, in our case whether or not a subject solves a psychological task.

\begin{figure}[ht]
    \centering
    \includegraphics[width=\textwidth]{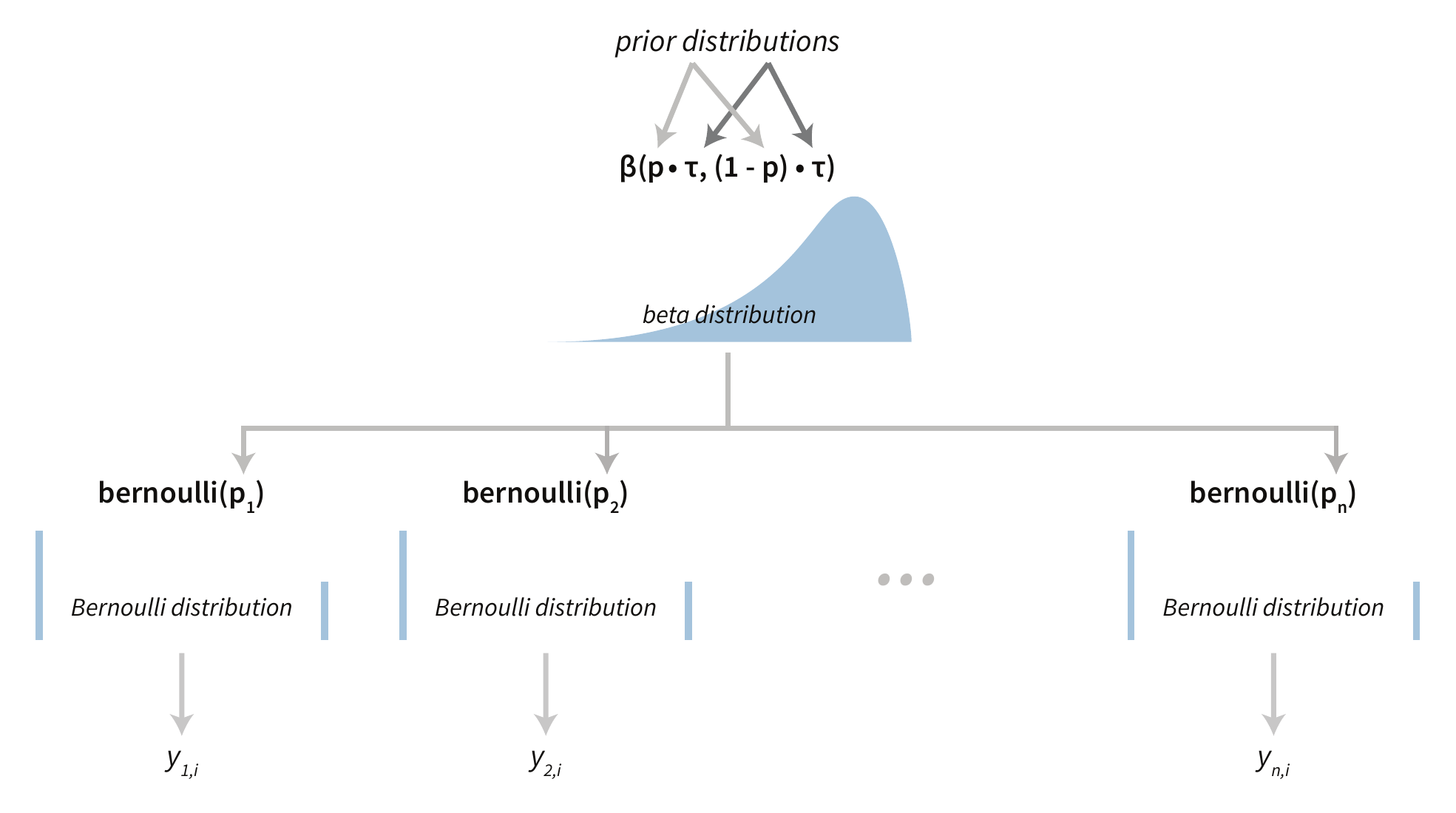}
    \caption{{\bf Visualization of the Bayesian success rate model.} The model has a hierarchical structure, data about success of individual subjects ($y_{n,i}$ depicts success on the $i$-th attempt of the subject $n$) is used for fitting Bernoulli distributions on the subject level. Parameters of subject level distributions are then connected at the group level with a Beta distribution.}
    \label{fig:success_rate}
\end{figure}

The success rates model also has a hierarchical structure. The success rate of individual subjects is modelled using Bernoulli distributions, where the $p_i$ is the success rate of subject $i$. A reparametrized Beta distribution, $\BE(p\tau, (1-p)\tau)$, is used as a hierarchical prior on subject-level parameters, where $p$ is the group level success rate and $\tau$ is the scale parameter. A graphical representation of our hierarchical success rate model can be seen in \figurename~\ref{fig:success_rate}. For a practical application of this model see Section \ref{sec:flanker}.

\subsection{Model for analysis of sequential tasks}

In some psychological experiments the data have a time component or are ordered in some other way. For example, when subjects participate in a sequence of questions or tasks. To model how a subject's performance changes over time, we implemented a hierarchical linear normal model.

The sequence for each individual subject is modelled using a simple linear model with subject-specific slope and intercept. To model the data at the group level we put hierarchical normal priors on all parameters of the subject-level linear models. The parameters of subject $i$ are thus $\alpha_i$ for the intercept and $\beta_i$ for the slope of the linear model along with $\sigma_i$ for modelling errors of the fit (residuals). And hierarchical normal priors on these parameters are $\mathcal{N}(\mu_\alpha, \sigma_\alpha)$ for the intercept ($\alpha$), $\mathcal{N}(\mu_\beta, \sigma_\beta)$ for the slope ($\beta$) parameter, along with $\mathcal{N}(\mu_\sigma, \sigma_\sigma)$ for the residuals ($\sigma$).

\begin{figure}[ht]
    \centering
    \includegraphics[width=\textwidth]{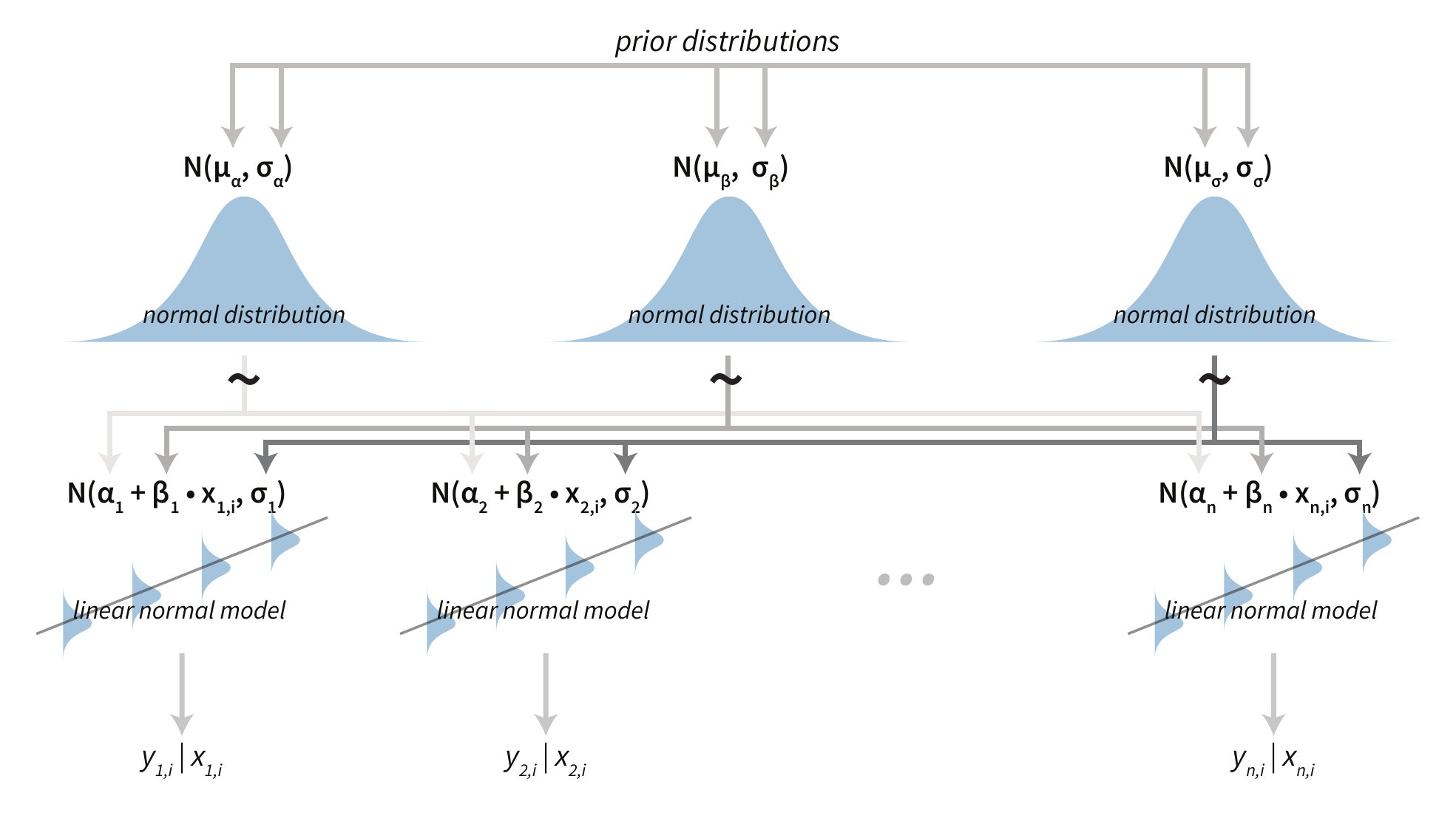}
    \caption{{\bf Visualization of the hierarchical linear model.} The model has a hierarchical structure, linear normal models are fitted on the subject level from data belonging to each particular subject. Since the ordering of results is important input data come in pairs of dependent (e.g. result or answer) and independent variables (e.g. time or the question index), $y_{n,i} | x_{n,i}$ depicts the value of the $i$-th dependent variable given the value of the independent variable $i$ for the subject $n$. Parameters of subject level distributions are joined on the group level by using normal distributions. These distributions can then be used for group level analysis of the data.}
    \label{fig:linear}
\end{figure}

A graphical representation of the model is shown in \figurename~\ref{fig:linear}. For a practical application of this model see Section \ref{sec:adaptation}.

\subsection{Model for analysis of color based tasks}

Color stimuli and subject responses in psychological experiments are most commonly defined through the RGB color model. The name of the model comes from the initials of the three additive primary colors, red, green and blue. These colors are also the three components of the model, each component has a value from 1 to 255 which defines the presence of a particular color. Since defining and analyzing colors through the RGB model is not very user friendly or intuitive, our Bayesian model is capable of working with both the RGB and HSV color models. HSV (hue, saturation and value) is an alternative representation of the RGB model that is much easier to read and interpret for most human beings.

\begin{figure}[ht]
    \centering
    \includegraphics[width=\textwidth]{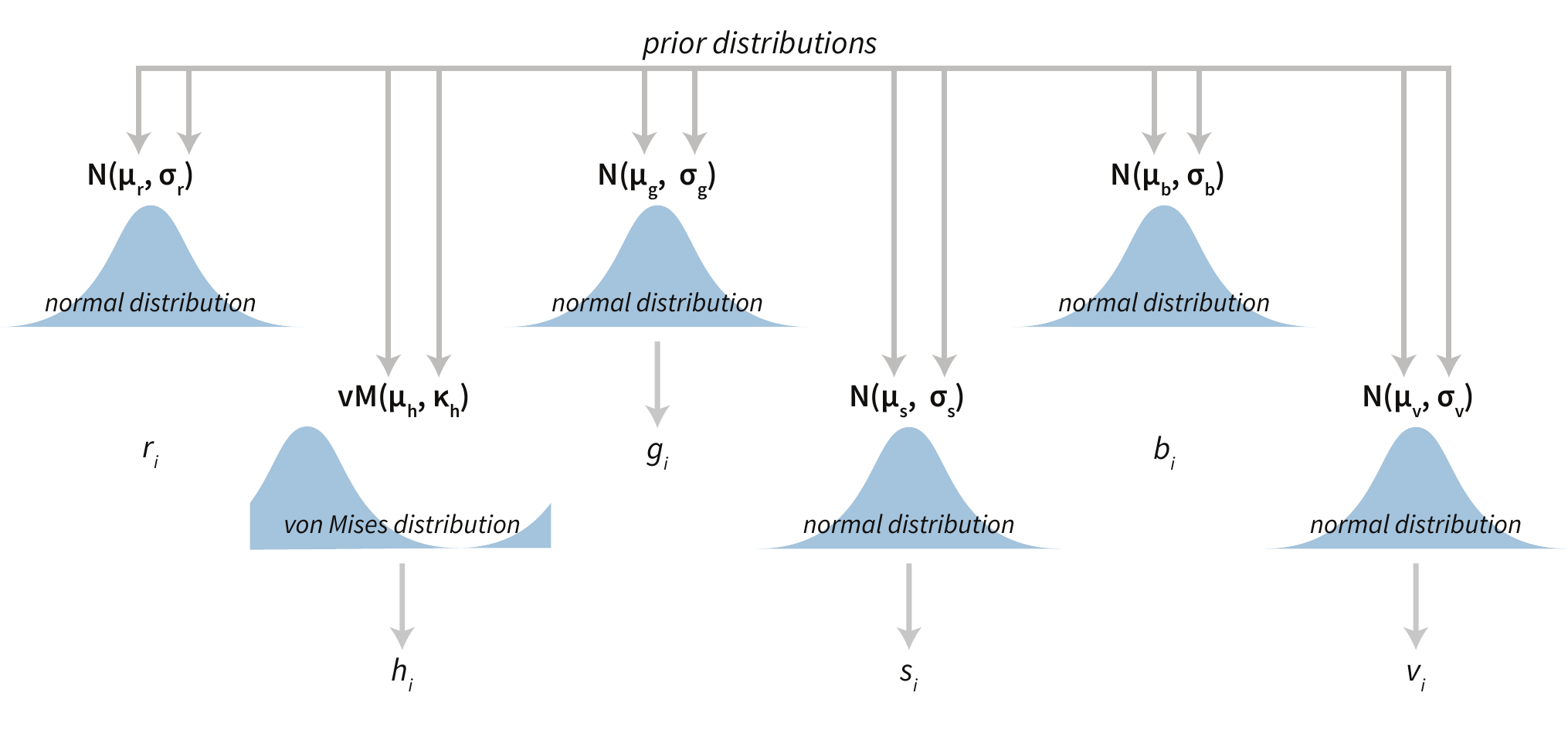}
    \caption{{\bf Visualization of the Bayesian color model.} The model is composed of six parts. Three parts are used to describe the RGB (red, green, blue) color model components and three parts are used to describe the HSV (hue, saturation, value) color model components. All components, except hue, are modeled with normal distributions, while hue is modelled with the von Mises distribution -- a circular normal distribution.}
    \label{fig:color}
\end{figure}

The Bayesian color model works in a component-wise fashion -- six distributions (three for the RGB components and three for the HSV components) are fitted to data for each component individually. For RGB components we use normal distributions (truncated to the [0, 255] interval). In the HSV case, we used normal distributions (truncated to the [0, 1] interval) for saturation and value components and the von Mises distribution for the hue component. The von Mises distribution (also known as the circular normal distribution) is a close approximation to the normal distribution wrapped on the [0, 2pi] interval. A visualization of our Bayesian model for colors can be seen in \figurename~\ref{fig:color} and its practical application in Section \ref{sec:afterimages}.

\subsection{Bayesian bootstrap}

Bootstrapping is a commonly used resampling technique for estimating statistics on a population by sampling a data set with replacement. It is usually used to for evaluating measures of accuracy (such as the mean, bias, standard deviation, confidence intervals ...) to sample estimates. It allows estimation of the sampling distribution of almost any statistic using random sampling methods and is as such useful in a wide repertoire of scenarios.

The Bayesian bootstrap inside the \pkg{bayes4psy} package is the analogue of the classical bootstrap \citep{Efron1979Bootstrap}. It is based on Rasmus Bååth's implementation \citep{Baath2016Bootstrap}, which in turn is based on methods developed by \citep{Rubin1981Bootstrap}. Bayesian bootstrap does not simulate the sampling distribution of a statistic estimating a parameter, but instead simulates the posterior distribution of the parameter. The statistical model underlying the Bayesian bootstrap can be characterized by drawing weights from a uniform Dirichlet distribution with the same dimension as the number of data points, these draws are then used for calculating the statistic in question and weighing the data \citep{Baath2016Bootstrap}. For more details about the implementation see \citep{Baath2016Bootstrap} and \citep{Rubin1981Bootstrap}.

\subsection{Methods for fitting and analyzing Bayesian fits}

This section provides a quick overview of all the methods for fitting and analyzing the models described in previous sections. For a more detailed description of each function readers are invited to consult the documentation and examples that are provided along with the \proglang{R} package.

The first set of functions constructs a Bayesian model from input data. Users can also use these functions to define priors (for an example, see the second part of the Section~\ref{sec:flanker}), set the number of MCMC iterations along with several other parameters of the MCMC algorithm (some basic MCMC settings are described in the documentation of this package, for more advanced settings consult the official \proglang{Stan} \citep{Carpenter2017Stan} documentation).

\begin{itemize}
    \item \code{b\_ttest} is a function for fitting the Bayesian t-test model, the input data is a vector of normally distributed real numbers.
    \item \code{b\_linear} to construct the hierarchical linear model for analyzing sequential tasks users have to provide three data vectors -- $x$ a vector containing values of the in dependant variable (time, question index ...), $y$ a vector containing values of the dependant variable (subject's responses) and $s$ a vector containing ids of subjects used for denoting that $x_i$/$y_i$ pair belongs to subject $i$. 
    \item \code{b\_reaction\_time} input data to the Bayesian reaction time model consists of two vectors -- vector $t$ includes reaction times while vector $s$ is used for linking reaction times with subjects.
    \item \code{b\_success\_rate} to fit the Bayesian success rate model users have to provide two data vectors. The first vector $r$ contains results of an experiment with binary outcomes (e.g. success/fail, hit/miss ...) and the second vector $s$ is used to link results to subjects.
    \item \code{b\_color} input data to this model is a three column matrix or a \code{data.frame} where each column represents one of the components of the chosen color model (RGB or HSV). If the input data are provided in the HSV format then users also have to set the $hsv$ parameter to \code{TRUE}.
    \item \code{b\_bootstrap} the mandatory input into the Bayesian bootstrap function is the input data and the statistics function. The input data can be in the form of a vector, matrix or a \code{data.frame}.
\end{itemize}

Before interpreting the results, users can use the following functions to check if the models provide a good representation of the data:

\begin{itemize}
    \item \code{plot\_trace} draws the Markov chain trace plot for main parameters of the model, providing a visual way to inspect sampling behavior and assess mixing across chains and convergence.
    \item \code{plot\_fit} draws the fitted distribution against the input data. With hierarchical models we can use the subjects parameter to draw fits on the subject level.
    \item \code{plot\_fit\_hsv} a special function for inspecting the fit of the color model by using a color wheel like visualization of HSV components.
\end{itemize}

For a summary of the posterior with Monte Carlo standard errors and confidence intervals users can use the \code{summary} or \code{print}/\code{show} functions.

\begin{itemize}
    \item \code{summary} prints summary statistics of the main model's parameters.
    \item \code{print}, \code{show} prints a more detailed summary of the model's parameters. It includes estimated means, Monte Carlo standard errors (\code{se\_mean}), confidence intervals, effective sample size (\code{n\_eff}, a crude measure of effective sample size), and the R-hat statistic for measuring auto-correlation. R-hat measures the potential scale reduction factor on split chains and equals 1 at convergence \citep{Brooks1998General,Gelman1992Inference}.
\end{itemize}

Users can also extract samples from the posterior for further analysis:

\begin{itemize}
    \item \code{get\_parameters} returns a \code{data.frame} of model's parameters. In hierarchical models this returns a \code{data.frame} of group level parameters.
    \item \code{get\_subject\_parameters} can be used to extract subject level parameters from hierarchical models.
\end{itemize}

The \code{compare\_means} function can be used for comparison of parameters that represent means of the fitted models, to visualize these means one can use the \code{plot\_means} function and for visualizing the difference between means the {plot\_means\_difference} function. All comparison functions (functions that print or visualize difference between fitted models) also offer the option of defining the region of practical equivalence (rope) by setting the \code{rope} parameter.

\begin{itemize}
    \item \code{compare\_means} prints and returns a \code{data.frame} containing the comparison results, can be used for comparing two or multiple models at the same time.
    \item \code{plot\_means\_difference} visualizes the difference of means between two or multiple models at the same time.
    \item \code{plot\_means} plots the distribution of parameters that depict means, can be used on a single or multiple models at the same time.
    \item \code{plot\_means\_hsv} a special function for the Bayesian color model that plots means of HSV components by using a color wheel like visualization.
\end{itemize}

The following set of functions works in a similar fashion as the one for comparing means, the difference is that this one compares entire distributions and not just the means. The comparison is based on drawing a large amount of samples from the distributions.

\begin{itemize}
    \item \code{compare\_distributions} prints and returns a \code{data.frame} containing the comparison results, can be used for comparing two or multiple models at the same time.
    \item \code{plot\_distributions\_difference} visualizes the difference of distributions underlying two or multiple fits at the same time.
    \item \code{plot\_distribution} plots the distributions underlying the fitted models, can be used on a single or multiple models at the same time.
    \item \code{plot\_distributions\_hsv} a special function for the Bayesian color model that plots the distribution behind HSV components by using a color wheel like visualization.
\end{itemize}

\section{Illustrations}
\label{sec:illustrations}

Below are illustrative practical examples of models in the \pkg{bayes4psy} package. Additional examples can be found in the online repository (\url{https://github.com/bstatcomp/bayes4psy_tools}). 

For the sake of brevity, we omitted similar visualizations and outputs, e.g. we provided the diagnostic outputs and visualizations only the first time they appered and omitted them later due to similarity.

\subsection{The flanker task}
\label{sec:flanker}

In the Eriksen flanker task \citep{Eriksen1974Flanker} participants are presented with an image of an odd number of arrows (usually five or seven). Their task is to indicate the orientation (left or right) of the middle arrow as quickly as possible whilst ignoring the flanking arrows on left and right. There are two types of stimuli in the task: in the \emph{congruent} condition (e.g. ‘<<<<<<<‘) both the middle arrow and the flanking arrows point in the same direction, whereas in the \emph{incongruent} condition (e.g. ‘<<<><<<‘) the middle arrow points to the opposite direction of the flanking arrows.

As the participants have to consciously ignore and inhibit the misleading information provided by the flanking arrows in the incongruent condition, the performance in the incongruent condition is robustly worse than in the congruent condition, both in terms of longer reaction times as well as higher proportion of errors. The difference between reaction times and error rates in congruent and incongruent cases is a measure of subject's ability to focus his or her attention and inhibit distracting stimuli.

In the illustration below we compare reaction times and error rates when solving the flanker task between the control group (healthy subjects) and the test group (subjects suffering from a certain medical condition).

First, we load package \pkg{bayes4psy} and package \pkg{dplyr} for data wrangling. Second, we load the data and split them into control and test groups. For reaction time analysis we use only data where the response to the stimuli was correct:

\begin{CodeChunk}
    \begin{CodeInput}
        R> library(bayes4psy)
        R> library(dplyr)
    
        R> data <- read.table("./data/flanker.csv",
                              sep="\t", header=TRUE)
        
        R> control_rt <- data %>% filter(result == "correct" &
                                         group == "control")
                                         
        R> test_rt <- data %>% filter(result == "correct" &
                                      group == "test")
    \end{CodeInput}
\end{CodeChunk}

The model requires subjects to be indexed from $1$ to $n$. Control group subject indexes range from 22 to 45, so we cast it to 1 to 23.

\begin{CodeChunk}
    \begin{CodeInput}
        R> control_rt$subject <- control_rt$subject - 21
    \end{CodeInput}
\end{CodeChunk}

Now we are ready to fit the Bayesian reaction time model for both groups. The model function requires two parameters --  a vector of reaction times $t$ and the vector of subject indexes $s$.

\begin{CodeChunk}
    \begin{CodeInput}
        R> rt_control_fit <- b_reaction_time(t=control_rt$rt,
                                             s=control_rt$subject)
                                             
        R> rt_test_fit <- b_reaction_time(t=test_rt$rt,
                                          s=test_rt$subject)                     
    \end{CodeInput}
\end{CodeChunk}

Before we interpret the results, we check MCMC diagnostics and model fit.

\begin{CodeChunk}
    \begin{CodeInput}
        plot_trace(rt_control_fit)
        plot_trace(rt_test_fit)
    \end{CodeInput}
\end{CodeChunk}

\clearpage

\begin{figure}[ht]
    \centering
    \includegraphics[width=\textwidth]{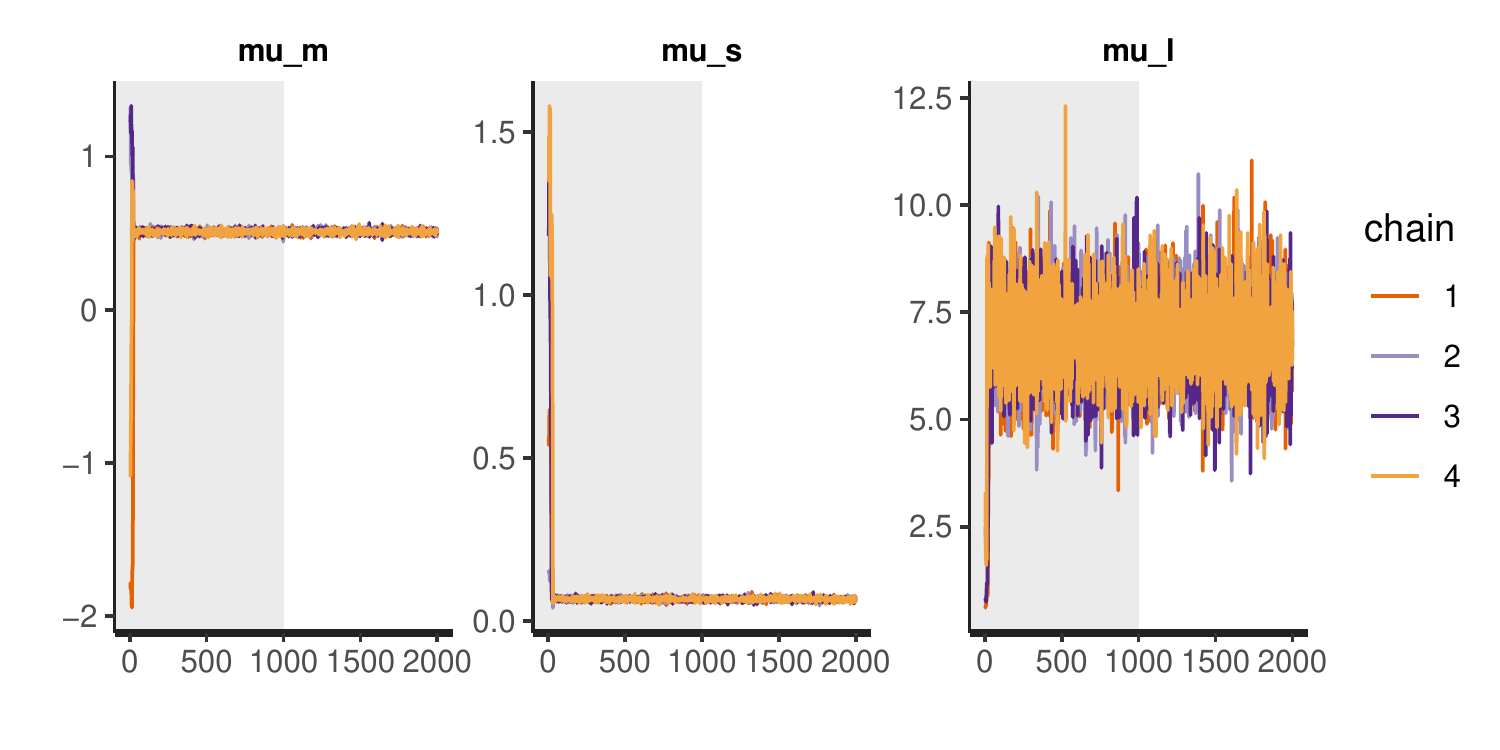}
    \caption{{\bf The trace plot for \code{rt\_control\_fit}.} The traceplot gives us no cause for concern regarding MCMC convergence and mixing. The trace plot for \code{rt\_test\_fit} is similar. Note that the first 1000 iterations (shaded gray) are used for warmup (tuning of the MCMC algorithm) and are discarded. The next 1000 iterations are used for sampling.}
    \label{fig:flanker_plot_trace}
\end{figure}

\begin{CodeChunk}
    \begin{CodeInput}
        R> print(rt_control_fit)
        
        Inference for Stan model: reaction_time.
        4 chains, each with iter=2000; warmup=1000; thin=1; 
        post-warmup draws per chain=1000, total post-warmup draws=4000.
        
                           mean se_mean   sd    2.5%    97.5% n_eff Rhat
        mu[1]              0.46    0.00 0.01    0.44     0.47  4789    1
        mu[2]              0.36    0.00 0.01    0.35     0.38  4661    1
        ...
        sigma[1]           0.04    0.00 0.01    0.03     0.05  5406    1
        sigma[2]           0.03    0.00 0.01    0.02     0.04  5165    1
        ...
        lambda[1]         14.41    0.02 1.62   11.59    17.87  4441    1
        lambda[2]         11.59    0.02 1.15    9.53    14.01  5271    1
        ...
        mu_m               0.51    0.00 0.01    0.48     0.54  5589    1
        mu_l               6.86    0.01 0.91    5.12     8.75  5299    1
        mu_s               0.07    0.00 0.01    0.06     0.08  4115    1
        sigma_m            0.06    0.00 0.01    0.05     0.09  6078    1
        sigma_l            4.24    0.01 0.78    3.02     5.99  3940    1
        sigma_s            0.02    0.00 0.00    0.01     0.03  3862    1
        rt                 0.66    0.00 0.02    0.61     0.71  5112    1
        rt_subjects[1]     0.53    0.00 0.01    0.51     0.54  4261    1
        rt_subjects[2]     0.45    0.00 0.01    0.44     0.47  5654    1
        ...
        
        R> print(rt_test_fit)
    \end{CodeInput}
\end{CodeChunk}

The output above is truncated and shows only values for 2 of the 24 subjects on the subject level of the hierarchical model. The output provides further MCMC diagnostics, which again do not give us cause for concern. The convergence diagnostic \code{Rhat} is practically 1 for all parameters and there is little auto-correlation (possibly even some positive auto-correlation) -- effective sample sizes (\code{n\_eff}) are of the order of samples taken and Monte Carlo standard errors (\code{se\_mean}) are relatively small.

What is a good-enough effective sample sizes depends on our goal. If we are interested in posterior quantities such as the more extreme percentiles, the effective sample sizes should be 10,000 or higher, if possible. If we are only interested in estimating the mean, 100 effective samples is in most cases enough for a practically negligible Monte Carlo error.

We can increase the effective sample size by increasing the amount of MCMC iterations with the \code{iter} parameter. In our case we can achieve an effective sample size of 10,000 by setting \code{iter} to 4000. Because the MCMC diagnostics give us no reason for concern, we can leave the \code{warmup} parameter at its default value of 1000.

\begin{CodeChunk}
    \begin{CodeInput}
        R> rt_control_fit <- b_reaction_time(t=control_rt$rt,
                                             s=control_rt$subject,
                                             iter=4000)
                                             
        R> rt_test_fit <- b_reaction_time(t=test_rt$rt,
                                          s=test_rt$subject,
                                          iter=4000)                     
    \end{CodeInput}
\end{CodeChunk}

Because we did not explicitly define any priors, flat (improper) priors were put on all of the model's parameters. In some cases, flat priors are a statement that we have no prior knowledge about the experiment results (in some sense). In general, even flat priors can express a preference for a certain region of parameter space. In practice, we will almost always have some prior information and we should incorporate it into the modelling process.

Next, we check whether the model fits the data well by using the \code{plot\_fit} function. If we set the \code{subjects} parameter to \code{FALSE}, we will get a less detailed group level fit.

\clearpage

\begin{CodeChunk}
    \begin{CodeInput}
        R> plot_fit(rt_control_fit)
        R> plot_fit(rt_test_fit)
    \end{CodeInput}
\end{CodeChunk}

\begin{figure}[ht]
    \centering
    \includegraphics[width=\textwidth]{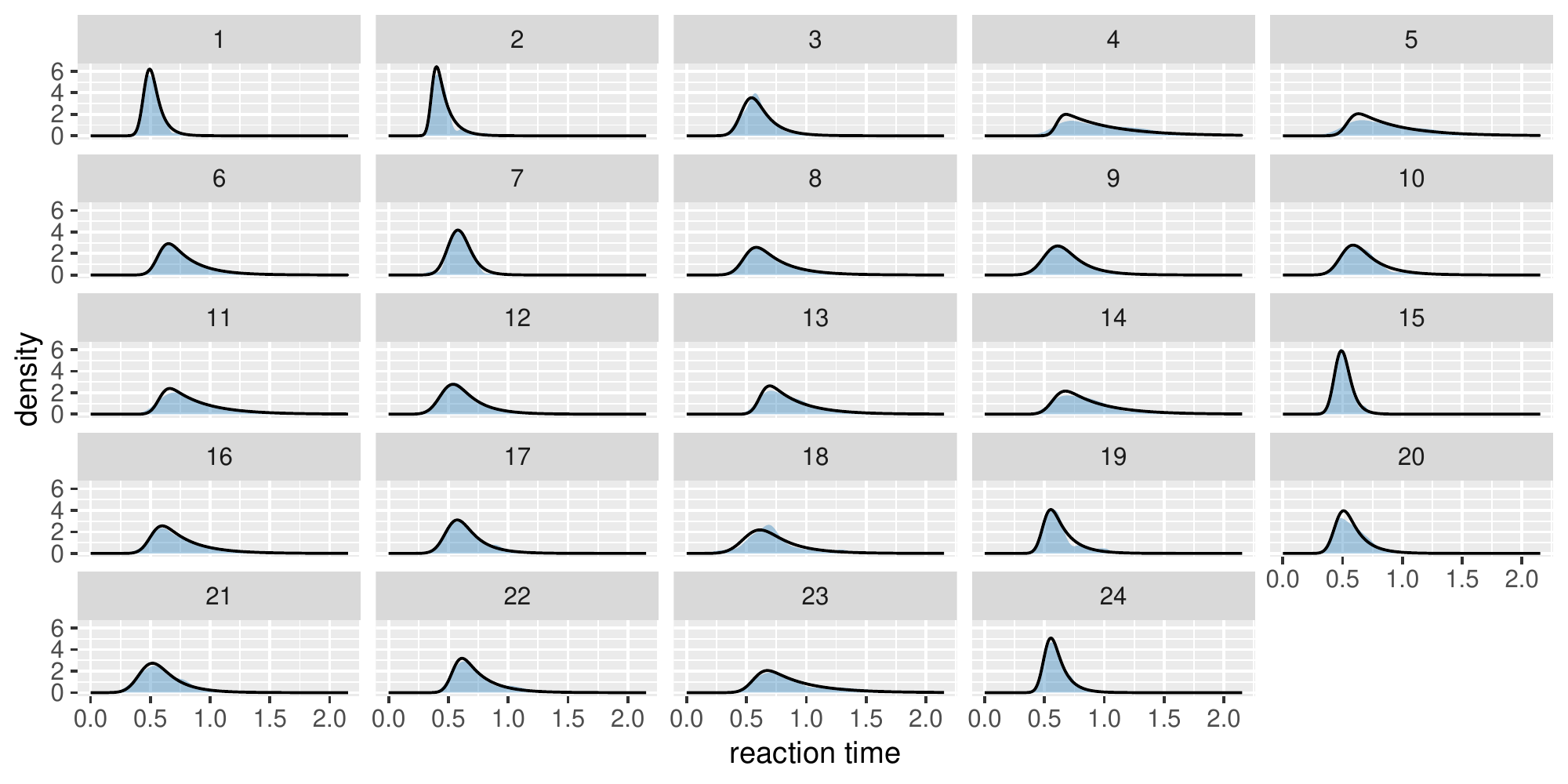}
    \caption{{\bf The fit plot for the \code{rt\_control\_fit}.} The data are visualized as a blue region while the fit is visualized with a black line. In this case the model fits the underlying data well, similar conclusions can be reached for the test group (\code{rt\_test\_fit}).}
    \label{fig:flanker_plot_fit}
\end{figure}

We now use the \code{compare\_means} function to compare reaction times between healthy (control) and unhealthy (test) subjects. In the example below we use a rope (region of practical equivalence) interval of 0.01 s, meaning that differences smaller that 1/100 of a second are deemed as equal. The \code{compare\_means} function provides a user friendly output of the comparison and also returns the results in the form of a \code{data.frame}.

\begin{CodeChunk}
    \begin{CodeInput}
        R> rt_control_test <- compare_means(rt_control_fit,
                                            fit2=rt_test_fit,
                                            rope=0.01)
        
        ---------- Group 1 vs Group 2 ----------
        Probabilities:
          - Group 1 < Group 2: 0.98 +/- 0.00409
          - Group 1 > Group 2: 0.01 +/- 0.00304
          - Equal: 0.01 +/- 0.00239
        95% HDI:
          - Group 1 - Group 2: [-0.17, -0.01]
    \end{CodeInput}
\end{CodeChunk}

The \code{compare\_means} function output contains probabilities that one group has shorter reaction times than the other, the probability that both groups are equal (if rope interval is provided) and the 95\% HDI (highest density interval, \cite{Kruschke2014BayesianDataAnalysis}) for the difference between groups. Based on the output we are quite certain (98\% +/- 0.5\%) that the healthy group's (\code{rt\_control\_fit}) expected reaction times are lower than the unhealthy group's (\code{rt\_test\_fit}).

We can also visualize this difference by using the \code{plot\_means\_difference} function. The \code{plot\_means} function is an alternative for comparing \code{rt\_control\_fit} and \code{rt\_test\_fit} -- the function visualizes the parameters that determine the means of each model.

\begin{CodeChunk}
    \begin{CodeInput}
        R> plot_means_difference(rt_control_fit,
                                 fit2=rt_test_fit,
                                 rope=0.01)                    
    \end{CodeInput}
\end{CodeChunk}

\begin{figure}[ht]
    \centering
    \includegraphics[width=\textwidth]{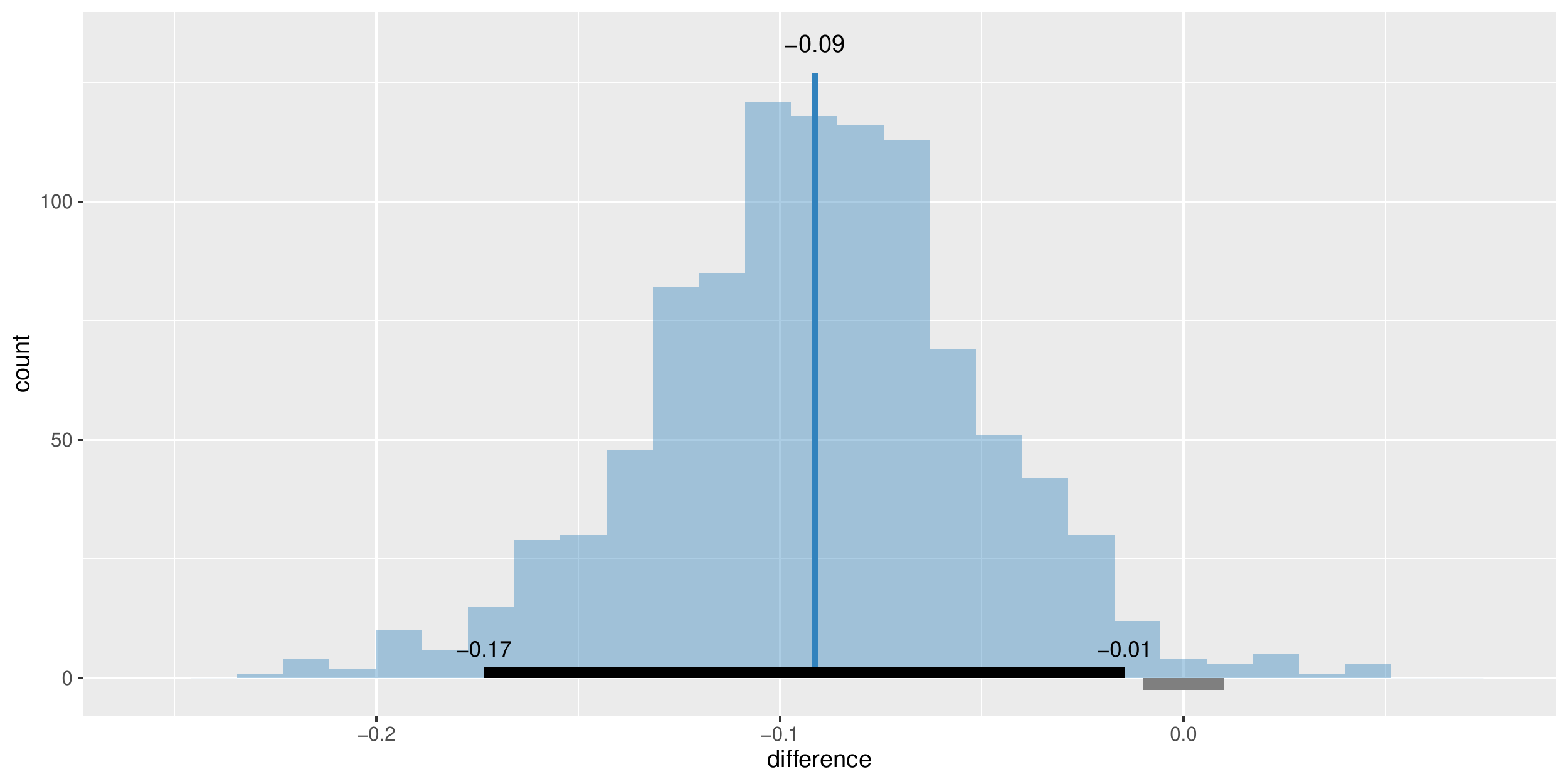}
    \caption{{\bf A visualization of the difference between \code{rt\_control\_fit} and \code{rt\_test\_fit}.} The histogram visualizes the distribution of the difference, vertical blue line denotes the mean, the black band at the bottom marks the 95\% HDI interval and the gray band marks the rope interval. Since the whole 95\% HDI of difference is negative and lies outside of the rope interval we can conclude that the statement that healthy subjects are faster than unhealthy ones is most likely correct.}
    \label{fig:flanker_plot_means_difference_rt}
\end{figure}

\clearpage

\begin{CodeChunk}
    \begin{CodeInput}
        R> plot_means(rt_control_fit,
                      fit2=rt_test_fit)                        
    \end{CodeInput}
\end{CodeChunk}

\begin{figure}[ht]
    \centering
    \includegraphics[width=\textwidth]{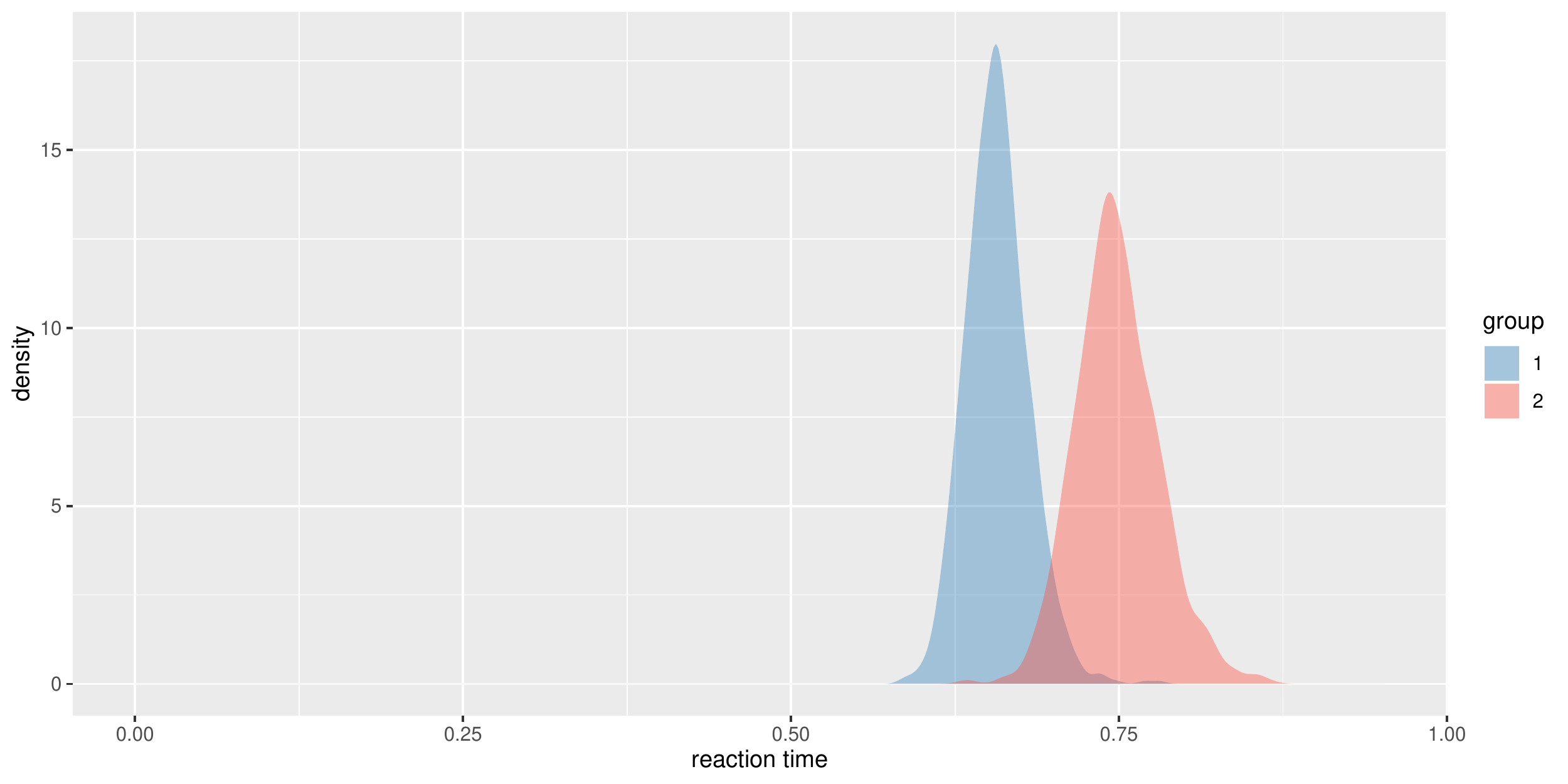}
    \caption{{\bf A visualization of means for \code{rt\_control\_fit} and \code{rt\_test\_fit}.} Group 1 visualizes means for the healthy subjects and group 2 for the unhealthy subjects.}
    \label{fig:flanker_plot_means}
\end{figure}

We can with high probability (98\% +/- 0.5\%) claim that healthy subjects have faster reaction times when solving the flanker task than unhealthy subjects. Next, we analyze if the same applies to success rates.

The information about success of subject's is stored as "correct"/"incorrect". However, the Bayesian success rate model requires binary inputs (0/1) so we have to transform the data. Also, like in the reaction time example, we have to correct the indexes of control group subjects.

\begin{CodeChunk}
    \begin{CodeInput}
        R> data$result_numeric <- 0
        R> data[data$result == "correct", ]$result_numeric <- 1

        R> control_sr <- data %>% filter(group == "control")
        R> test_sr <- data %>% filter(group == "test")

        R> control_sr$subject <- control_sr$subject - 21
    \end{CodeInput}
\end{CodeChunk}

Since the only prior information about the success rate of participants was the fact that success rate is located between 0 and 1, we used a beta distribution to put a uniform prior on the [0, 1] interval (we put a $\BE(1, 1)$ prior on the $p$ parameter). We execute the model fitting by running the \code{b\_success\_rate} function with appropriate input data.

\begin{CodeChunk}
    \begin{CodeInput}
        R> p_prior <- b_prior(family="beta", pars=c(1, 1))
        
        R> priors <- list(c("p", p_prior))
        
        R> sr_control_fit <- b_success_rate(r=control_sr$result_numeric,
                                            s=control_sr$subject,
                                            priors=priors,
                                            iter=4000)
        
        R> sr_test_fit <- b_success_rate(r=test_sr$result_numeric,
                                         s=test_sr$subject,
                                         priors=priors,
                                         iter=4000)
    \end{CodeInput}
\end{CodeChunk}

The process for inspecting Bayesian fits is the same and the results are similar as above, so we omit them. When visually inspecting the quality of the fit (the \code{plot\_fit} function) we can set the \code{subjects} parameter to \code{FALSE}, which visualizes the fit on the group level. This offers a quicker, but less detailed method of inspection.

\begin{CodeChunk}
    \begin{CodeInput}
        R> plot_trace(sr_control_fit)
        R> plot_trace(sr_test_fit)

        R> print(sr_control_fit)
        R> print(sr_test_fit)

        R> plot_fit(sr_control_fit, subjects=FALSE)
        R> plot_fit(sr_test_fit, subjects=FALSE)
    \end{CodeInput}
\end{CodeChunk}

Since diagnostic functions show no pressing issues and the fits look good we can proceed with the actual comparison between the two fitted models. We will again estimate the difference between two groups with \code{compare\_means}.

\begin{CodeChunk}
    \begin{CodeInput}
        R> sr_control_test <- compare_means(sr_control_fit, fit2=sr_test_fit)

        ---------- Group 1 vs Group 2 ----------
        Probabilities:
          - Group 1 < Group 2: 0.53 +/- 0.01052
          - Group 1 > Group 2: 0.47 +/- 0.01052
        95% HDI:
          - Group 1 - Group 2: [-0.02, 0.02]
    \end{CodeInput}
\end{CodeChunk}

As we can see the success rate between the two groups is not that different. Since the probability that healthy group is more successful is only 53\% (+/- 1\%) and the 95\% HDI of the difference ([0.02, 0.02]) includes the 0 we cannot claim inequality \citep{Kruschke2014BayesianDataAnalysis}. We can visualize this result by using the \code{plot\_means\_difference} function.

\clearpage

\begin{CodeChunk}
    \begin{CodeInput}
        R> plot_means_difference(sr_control_fit, fit2=sr_test_fit)
    \end{CodeInput}
\end{CodeChunk}

\begin{figure}[ht]
    \centering
    \includegraphics[width=\textwidth]{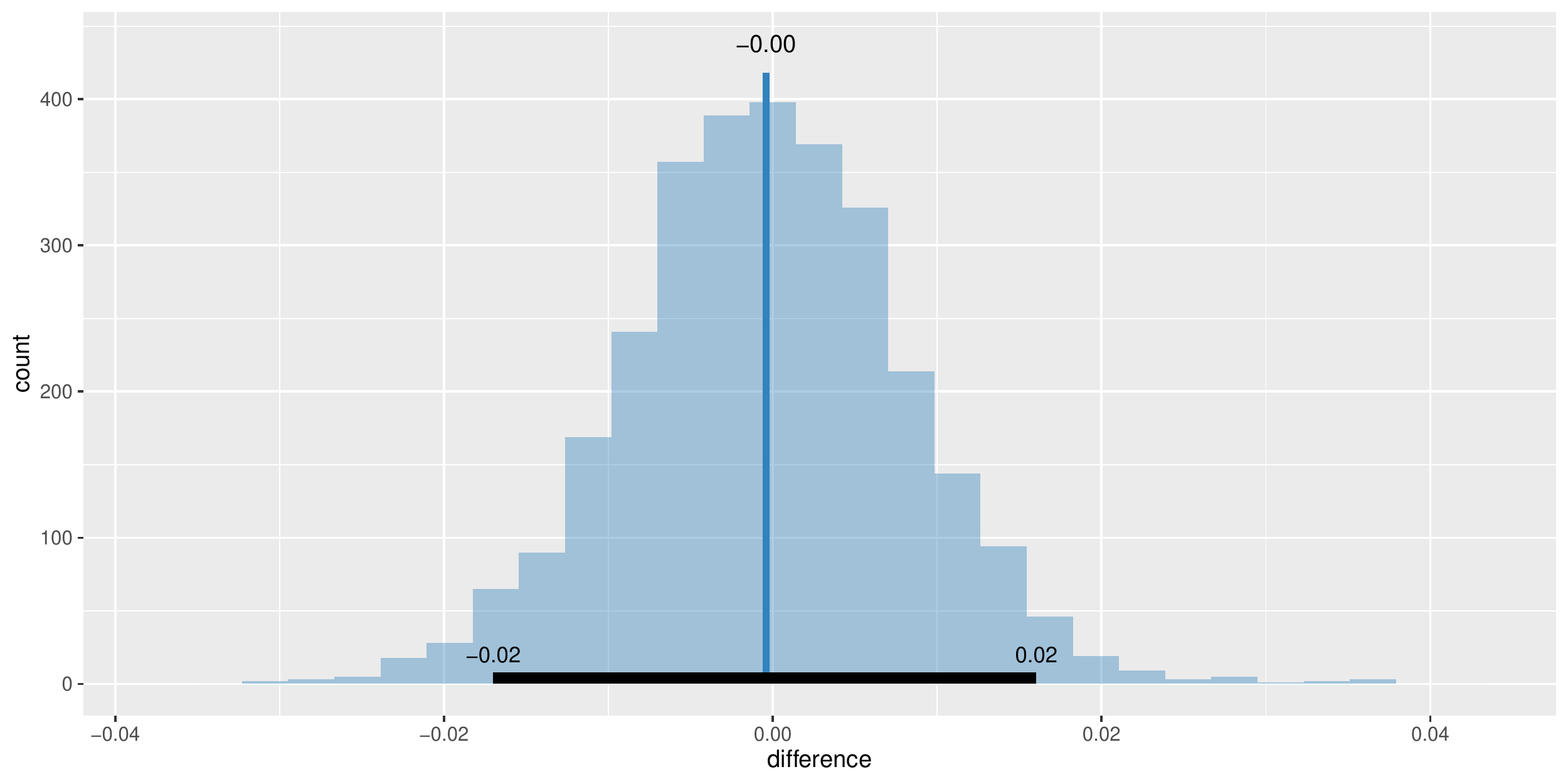}
    \caption{{\bf A visualization of the difference between the \code{sr\_control\_fit} and the \code{sr\_test\_fit}.} Histogram visualizes the distribution of the difference, vertical blue line denotes the mean difference and the black band at the bottom marks the 95\% HDI interval. Since the 95\% HDI of difference includes the value of 0 we cannot claim inequality. If we used a rope interval and the whole rope interval lied in the 95\% HDI interval we could claim equality.}
    \label{fig:flanker_plot_means_difference_sr}
\end{figure}

\clearpage

\subsection{Adaptation level}
\label{sec:adaptation}

In the adaptation level experiment participants had to assess weights of the objects placed in their hands by using a verbal scale: very very light, very light, light, medium light, medium, medium heavy, 
heavy, very heavy and very very heavy. The task was to assess the weight of an object that was placed on the palm of their hand. To standardize the procedure the participants had to place the elbow on the desk, extend the palm and assess the weight of the object after it was placed on their palm by slight up and down movements of their arm. During the experiment participants were blinded by using non-transparent fabric. In total there were 15 objects of the same shape and size but different mass (photo film canisters filled with metallic balls). Objects were grouped into three sets:

\begin{itemize}
    \item light set: 45 g, 55 g, 65 g, 75 g, 85 g (weights 1 to 5),
    \item medium set: 95 g, 105 g, 115 g, 125 g, 135 g (weights 6 to 10),
    \item heavy set: 145 g, 155 g, 165 g, 175 g, 185 g (weights 11 to 15).
\end{itemize}

The experimenter sequentially placed weights in the palm of the participant and recorded the trial index, the weight of the object and participant's response. The participants were divided into two groups, in group 1 the participants first assessed the weights of the light set in ten rounds within which all five weights were weighted in a random order. After completing the 10 rounds with the light set, the experimenter switched to the medium set, without any announcement or break. The participant then weighted the medium set across another 10 rounds of weighting the five weights in a random order within each round. In group 2 the overall procedure was the same, the only difference being that they started with the 10 rounds of the heavy set and then performed another 10 rounds of weighting of the medium set. Importantly, the weights within each set were given in random order and the experimenter switched between sets seamlessly without any break or other indication to the participant.

We will use the \pkg{bayes4psy} package to show that the two groups provide different assessment of the weights in the second part of the experiment even though both groups are responding to weights from the same (medium) set. The difference is very pronounced at first but then fades away with subsequent assessments of medium weights. This is congruent with the hypothesis that each group formed a different adaptation level during the initial phase of the task, the formed adaptation level then determined the perceptual experience of the same set of weights at the beginning of the second part of the task.

We will conduct the analysis by using the hierarchical linear model and the Bayesian t-test. First we have to construct fits for the second part of the experiment for each group independently. The code below loads and prepares the data, just like in the previous example, subject indexes have to be mapped to a [1, n] interval. We will use to \pkg{ggplot2} package to fine-tune graph axes and properly annotate graphs returned by the \pkg{bayes4psy} package.

\begin{CodeChunk}
    \begin{CodeInput}
        R> library(bayes4psy)
        R> library(dplyr)
        R> library(ggplot2)
        
        R> data <- read.table("./data/adaptation_level.csv",
                               sep="\t", header=TRUE)
                               
        R> group1 <- data %>% filter(group == 1)
        R> group2 <- data %>% filter(group == 2)
        
        R> n1 <- length(unique(group1$subject))
        R> n2 <- length(unique(group2$subject))
        
        R> group1$subject <- plyr::mapvalues(group1$subject,
                                             from=unique(group1$subject),
                                             to=1:n1)
        R> group2$subject <- plyr::mapvalues(group1$subject,
                                             from=unique(group1$subject),
                                             to=1:n2)
                                             
        R> group1_part2 <- group1 %>% filter(part == 2)
        R> group2_part2 <- group2 %>% filter(part == 2)
    \end{CodeInput}
\end{CodeChunk}

Once the data is prepared we can fit the Bayesian models, the input data comes in the form of three vectors, $x$ stores indexes of the measurements, $y$ subject's responses and $s$ indexes of subjects. The \code{warmup} and \code{iter} parameters are set in order to achieve an effective sample size of 10,000.

\begin{CodeChunk}
    \begin{CodeInput}
        R> fit1 <- b_linear(x=group1_part2$sequence,
                            y=group1_part2$response,
                            s=group1_part2$subject,
                            iter=10000, warmup=500)
        
        R> fit2 <- b_linear(x=group2_part2$sequence,
                            y=group2_part2$response,
                            s=group2_part2$subject,
                            iter=10000, warmup=500)
    \end{CodeInput}
\end{CodeChunk}

The fitting process is always followed by the quality analysis.

\begin{CodeChunk}
    \begin{CodeInput} 
        R> plot_trace(fit1)
        R> plot_trace(fit1)

        R> print(fit1)
        Inference for Stan model: linear.
        4 chains, each with iter=10000; warmup=500; thin=1; 
        post-warmup draws per chain=9500, total post-warmup draws=38000.
        
                     mean se_mean   sd    2.5%   97.5% n_eff Rhat
        alpha[1]     7.66    0.00 0.31    7.07    8.28 25452    1
        alpha[2]     8.63    0.00 0.23    8.19    9.08 23074    1
        ...
        beta[1]     -0.14    0.00 0.04   -0.24   -0.06 20097    1
        beta[2]     -0.12    0.00 0.03   -0.19   -0.05 30442    1
        ...
        sigma[1]     1.67    0.00 0.15    1.41    2.00 45998    1
        sigma[2]     0.99    0.00 0.10    0.82    1.21 44379    1
        ...
        mu_a         8.05    0.00 0.18    7.68    8.41 25983    1
        mu_b        -0.11    0.00 0.02   -0.15   -0.07 20126    1
        mu_s         1.10    0.00 0.09    0.92    1.29 33871    1
        sigma_a      0.61    0.00 0.16    0.38    0.98 24984    1
        sigma_b      0.05    0.00 0.02    0.01    0.09  6726    1
        sigma_s      0.34    0.00 0.08    0.21    0.54 30901    1
        lp__      -374.28    0.09 6.47 -387.21 -361.12  5372    1
        
        R> print(fit1_part2)
               
        R> plot_fit(fit1)
        R> plot_fit(fit1)    
    \end{CodeInput}
\end{CodeChunk}

The trace plot showed no MCMC related issues (for an example of trace plot see \figurename~\ref{fig:flanker_plot_trace}), effective sample sizes of parameters relevant for our analysis ($\mu_a$, $\mu_b$ and $\mu_s$) are large enough. Since the visual inspection of the fit also looks good we can continue with our analysis. To get a quick description of fits we can take a look at the summary statistics of model's parameters.

\begin{CodeChunk}
    \begin{CodeInput} 
        R> summary(fit1)
        intercept (alpha):  8.05 +/- 0.00266, 95% HDI: [7.69, 8.39]
        slope (beta):       -0.11 +/- 0.00033, 95% HDI: [-0.15, -0.07]
        sigma:              1.10 +/- 0.00094, 95% HDI: [0.91, 1.28]
        
        R> summary(fit2)
        intercept (alpha):  5.81 +/- 0.00461, 95% HDI: [5.20, 6.43]
        slope (beta):       0.12 +/- 0.00036, 95% HDI: [0.08, 0.16]
        sigma:              1.40 +/- 0.00165, 95% HDI: [1.13, 1.66]
    \end{CodeInput}
\end{CodeChunk}

Values of intercept suggest that our initial hypothesis about adaptation level is true. Subject's that weighted lighter object in the first part of the experiment (\code{fit1}) find medium objects at the beginning of experiment's second part heavier than subjects that weighted heavier objects in the first part (\code{fit2}). We can confirm this assumption by using functions that perform a more detailed analysis (e.g. \code{compare\_means} and \code{plot\_means\_difference}).

\begin{CodeChunk}
    \begin{CodeInput}
        R> comparison_results <- compare_means(fit1, fit2=fit2)
        ---------- Intercept ----------
        Probabilities:
          - Group 1 < Group 2: 0.00 +/- 0.00000
          - Group 1 > Group 2: 1.00 +/- 0.00000
        95% HDI:
          - Group 1 - Group 2: [1.54, 2.91]
        
        ---------- Slope ----------
        Probabilities:
          - Group 1 < Group 2: 1.00 +/- 0.00000
          - Group 1 > Group 2: 0.00 +/- 0.00000
        95% HDI:
          - Group 1 - Group 2: [-0.29, -0.18]

        R> plot_means_difference(fit1, fit2=fit2, par="intercept")
    \end{CodeInput}
\end{CodeChunk}

\begin{figure}[ht]
    \centering
    \includegraphics[width=\textwidth]{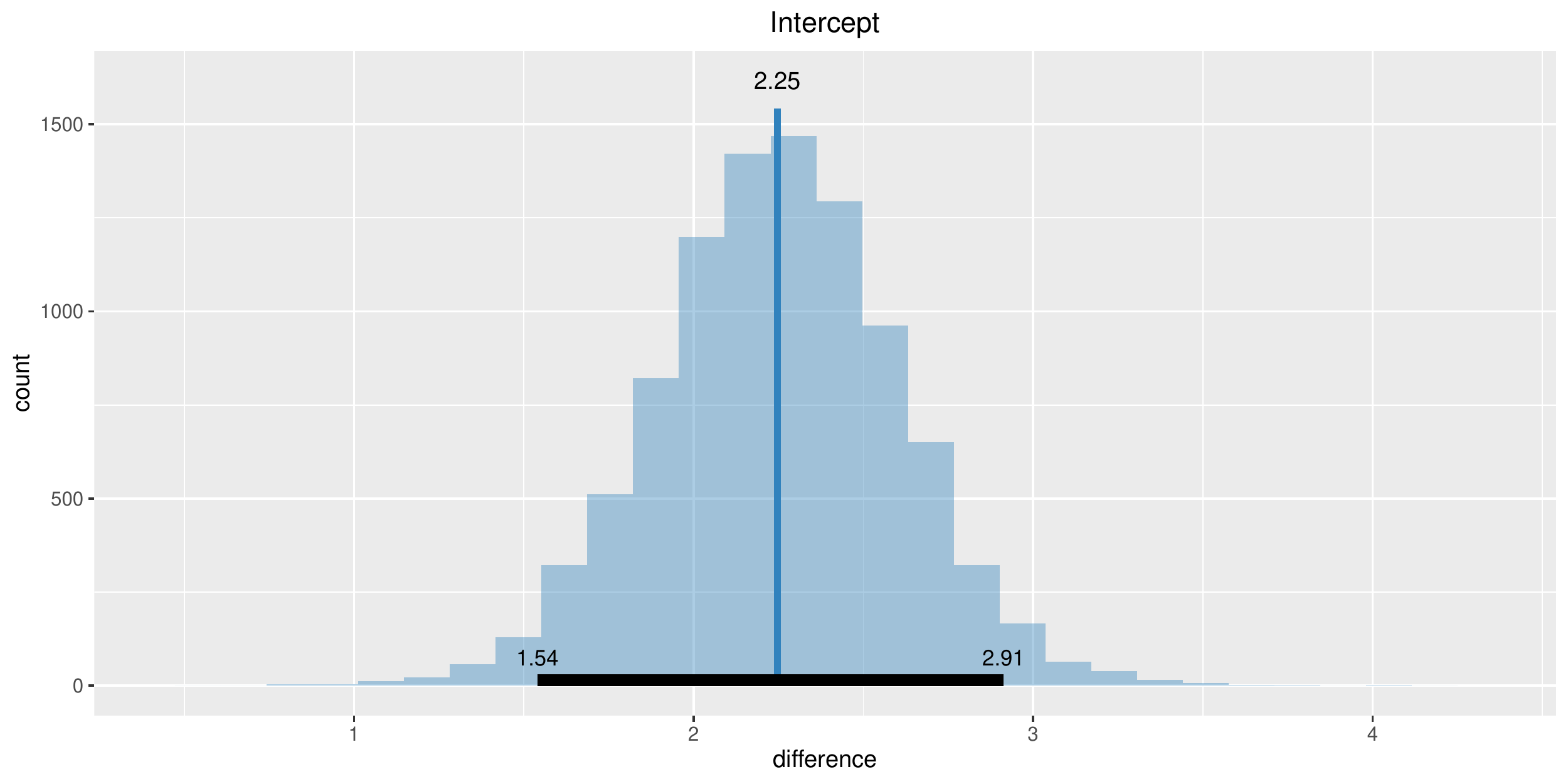}
    \caption{{\bf Difference of intercept between the two fits.} Since the whole 95\% HDI is positive we are quite confident that the subject's that weighted lighter object in the first part of the experiment (\code{fit1}) find medium objects heavier than subjects that initially weighted heavier objects (\code{fit2}).}
    \label{fig:al_plot_means_difference}
\end{figure}

The fact that the slope for the first group is very likely to be negative (the whole 95\% HDI lies below 0) and positive for the second group (the whole 95\% HDI lies above 0) suggests that the adaptation level phenomenon fades away with time. We can visualize this by plotting means and distributions underlying both fits. The plotting functions in the \pkg{bayes4psy} package return regular \pkg{ggplot2} plot objects, so we can use the same techniques to annotate or change the look and feel of graphs as we would with the usual \pkg{ggplot2} visualizations.

\clearpage

\begin{CodeChunk}
    \begin{CodeInput}
        R> plot_distributions(fit1, fit2) +
               labs(title="Part II", x="measurement number", y="") +
               theme(legend.position=) +
               theme(legend.position="none") +
               scale_x_continuous(limits=c(1, 10), breaks=seq(1:10)) +
               ylim(0, 10)
    \end{CodeInput}
\end{CodeChunk}

\begin{figure}[ht]
    \centering
    \includegraphics[width=\textwidth]{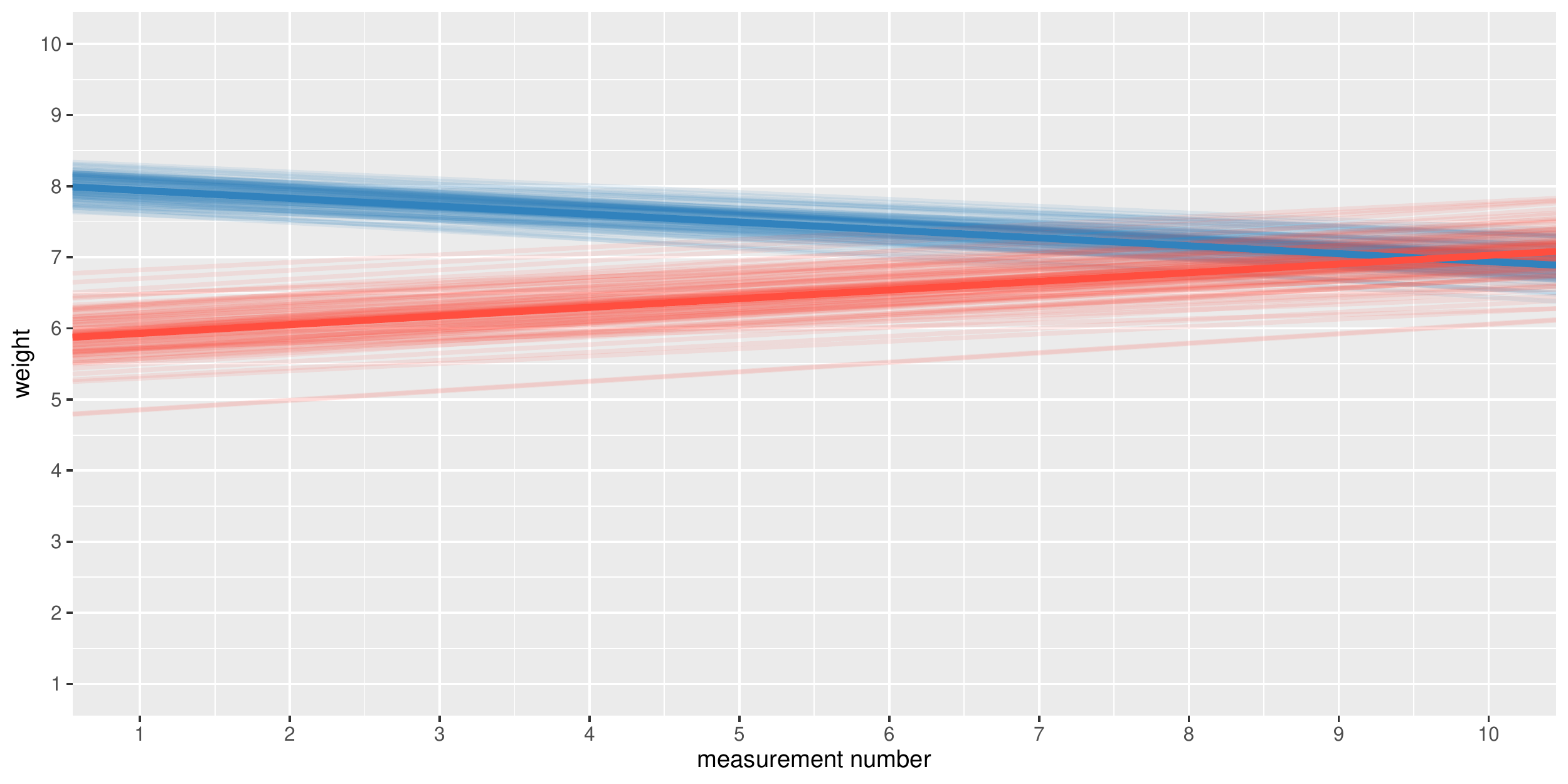}
    \caption{{\bf Comparison of distributions underlying \code{fit1} and \code{fit2}.} The hypothesis that each group formed a different adaptation level during the initial phase of the task seems to be true. Group that switches from heavy to medium weights assesses weights as lighter than they really are while for the group that switches from light to medium the weights appear heavier. With time these adaptation levels fade away and assessments converge to the same estimates of weight.}
    \label{fig:plot_distributions}
\end{figure}

\clearpage

\subsection{The Stroop color-word test}
\label{sec:stroop}
 
The Stroop test \citep{Stroop1935Studies} showed that when the stimuli is incongruent -- the name of a color is printed in different ink than the one denoted by its name (for example, \begin{color}{blue}red\end{color}) -- naming the color takes longer and is more error-prone than naming the color of a rectangle or a set of characters that does not form a word (for example, \begin{color}{blue}XXXXX\end{color})).

In our version of the Stroop test participants were faced with four types of conditions:

\begin{itemize}
    \item Reading neutral -- the name of the color was printed in black ink, the participant had to read the color's name.
    \item Naming neutral -- string XXXXX was written in colored ink (red, green or blue), the participant had to name the ink color.
    \item Reading incongruent -- name of the color was printed in incongruent ink, the participant had to read the written name of the color.
    \item Naming incongruent -- name of the color was printed in incongruent ink, the participant had to name the ink color.
\end{itemize}

We are primarily interested in expected task completion times. Every participant had the same number of stimuli in every condition, so we opt for a Bayesian t-test. The data are already split into the four conditions described above, so we only need to specify the priors. We based them on our previous experience with similar tasks -- participants finish the task in approximately 1 minute and the typical standard deviation for a participant is less than 2 minutes.

\begin{CodeChunk}
    \begin{CodeInput}
        R> library(bayes4psy)
        R> library(dplyr)
        R> library(ggplot2)
    
        R> data <- read.table("./data/stroop_simple.csv",
           sep="\t", header=TRUE)

        R> mu_prior <- b_prior(family="normal", pars=c(60, 30))
        R> sigma_prior <- b_prior(family="uniform", pars=c(0, 120))

        R> priors <- list(c("mu", mu_prior),
                          c("sigma", sigma_prior))

        R> fit_reading_neutral <- b_ttest(data$reading_neutral,
                                          priors=priors,
                                          iter=4000, warmup=500)
        R> fit_reading_incongruent <- b_ttest(data$reading_incongruent,
                                              priors=priors,
                                              iter=4000, warmup=500)
        R> fit_naming_neutral <- b_ttest(data$naming_neutral,
                                         priors=priors
                                         iter=4000, warmup=500)
        R> fit_naming_incongruent <- b_ttest(data$naming_incongruent,
                                             priors=priors,
                                             iter=4000, warmup=500)       
    \end{CodeInput}
\end{CodeChunk}

There were no reasons for concern in the MCMC diagnostics and model fits, so we omit them for brevity. In practice, we should of course always perform these steps.

We proceed by cross-comparing several fits with a single line of code.

\begin{CodeChunk}
    \begin{CodeInput}
        R> fit_list <- c(fit_reading_incongruent,
                         fit_naming_neutral,
                         fit_naming_incongruent)

        R> multiple_comparison <- compare_means(fit_reading_neutral,
                                                fits=fit_list)
        ---------- Group 1 vs Group 2 ----------
        Probabilities:
          - Group 1 < Group 2: 1.00 +/- 0.00054
          - Group 1 > Group 2: 0.00 +/- 0.00054
        95% HDI:
          - Group 1 - Group 2: [-4.66, -0.96]
                
        ---------- Group 1 vs Group 3 ----------
        Probabilities:
          - Group 1 < Group 3: 1.00 +/- 0.00000
          - Group 1 > Group 3: 0.00 +/- 0.00000
        95% HDI:
          - Group 1 - Group 2: [-15.34, -10.19]
              
        ---------- Group 1 vs Group 4 ----------
        Probabilities:
          - Group 1 < Group 4: 1.00 +/- 0.00000
          - Group 1 > Group 4: 0.00 +/- 0.00000
        95% HDI:
          - Group 1 - Group 2: [-36.72, -28.44]
              
        ---------- Group 2 vs Group 3 ----------
        Probabilities:
          - Group 2 < Group 3: 1.00 +/- 0.00000
          - Group 2 > Group 3: 0.00 +/- 0.00000
        95% HDI:
          - Group 1 - Group 2: [-12.63, -7.09]
               
        ---------- Group 2 vs Group 4 ----------
        Probabilities:
          - Group 2 < Group 4: 1.00 +/- 0.00000
          - Group 2 > Group 4: 0.00 +/- 0.00000
        95% HDI:
          - Group 1 - Group 2: [-34.12, -25.48]
               
        ---------- Group 3 vs Group 4 ----------
        Probabilities:
          - Group 3 < Group 4: 1.00 +/- 0.00000
          - Group 3 > Group 4: 0.00 +/- 0.00000
        95% HDI:
          - Group 1 - Group 2: [-24.21, -14.88]
        
        ----------------------------------------         
        Probabilities that a certain group is
        smallest/largest or equal to all others:
        
          largest     smallest equal
        1       0 0.9991111111     0
        2       0 0.0008888889     0
        3       0 0.0000000000     0
        4       1 0.0000000000     0
    \end{CodeInput}
\end{CodeChunk}

\begin{figure}[ht]
    \centering
    \includegraphics[width=\textwidth]{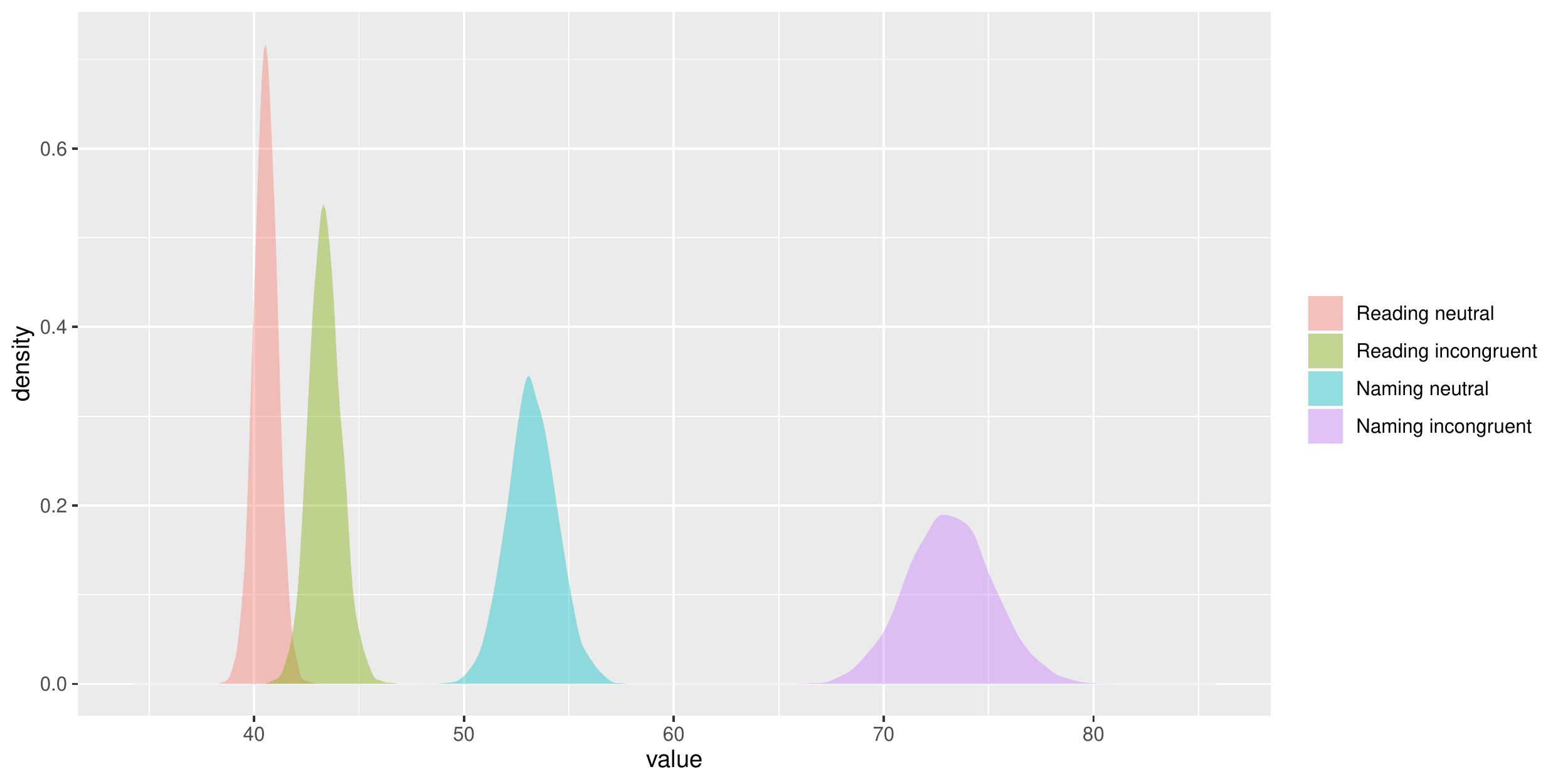}
    \caption{{\bf A visualization of means for all four types of Stroop tasks.} X-axis (value) denotes task completion time. Naming and incongruency conditions make the task more difficult, with naming having a bigger effect.}
    \label{fig:stroop_plot_means}
\end{figure}

\begin{figure}[ht]
    \centering
    \includegraphics[width=0.94\textwidth]{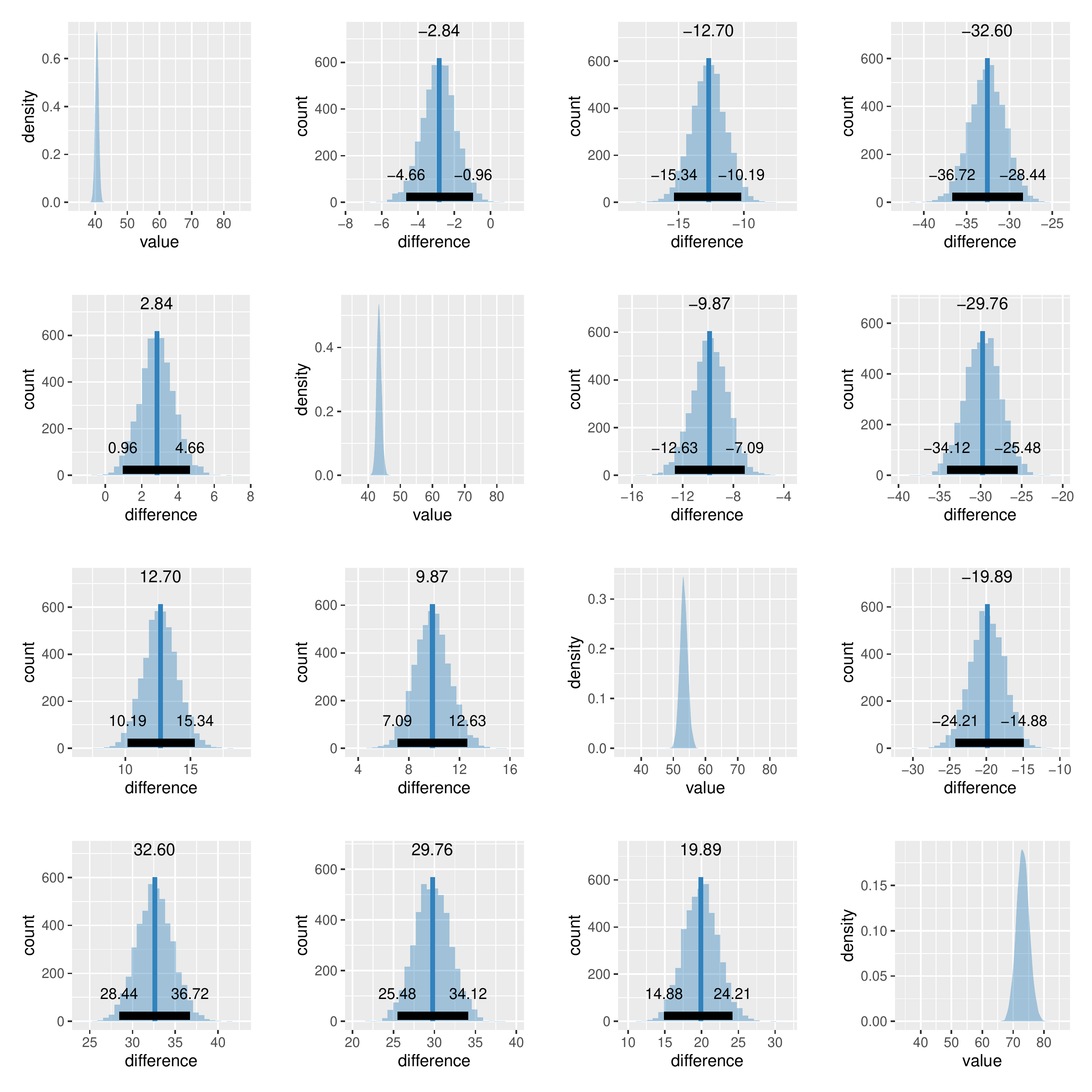}
    \caption{{\bf Differences in the mean task completion times for the four conditions..} Row and column 1 represent the reading neutral task, row and column 2 the reading incongruent task, row and column 3 the naming neutral task and row and column 4 the naming incongruent task. Since 95\% HDI intervals in all cases exclude 0 we are confident that the task completion times are different.}
    \label{fig:stroop_plot_means_difference}
\end{figure}

When we compare more than 2 fits, we also get an estimate of the probabilities that a group has the largest or the smallest expected value. Based on the above output, the participants are best at the reading neutral task (Group 1), followed by the reading incongruent task (Group 2) and the naming neutral task (Group 3). They are the worst at the naming incongruent task (Group 4). This ordering is true with very high probability, so we can conclude that both naming and incongruency of stimuli increase response times of subjects, with naming having a bigger effect. We can also visualize this in various ways, either as distributions of mean times needed to solve the given tasks (\figurename~\ref{fig:stroop_plot_means}) or as a difference between these means (\figurename~\ref{fig:stroop_plot_means_difference}).

\clearpage

\begin{CodeChunk}
    \begin{CodeInput}
        R> plot_means(fit_reading_neutral, fits=fit_list) +
            scale_fill_hue(labels=c("Reading neutral",
                                    "Reading incongruent",
                                    "Naming neutral",
                                    "Naming incongruent")) +
            theme(legend.title=element_blank())            
    \end{CodeInput}
\end{CodeChunk}

\clearpage

\begin{CodeChunk}
    \begin{CodeInput}
        R> plot_means_difference(fit_reading_neutral, fits=fit_list)            
    \end{CodeInput}
\end{CodeChunk}

\clearpage

\subsection{Afterimages}
\label{sec:afterimages}

In the afterimages task participants were asked to fix their gaze on a fixation point in the middle of the computer screen. Stimulus -- a colored rectangle -- was then shown above the fixation point. After 20 seconds the rectangle disappeared and a color palette was shown on the right-hand side of the screen. Participants were asked to keep their gaze on the fixation point while using the mouse to select the color that best matches the color of the afterimage that appeared above the fixation point. Then a colored rectangle of the selected color and same size as before was shown below the fixation point. Finally, participants confirmed their selection. For each trial the color of the stimulus rectangle, the response in RGB and the response time were recorded. The goal of this study was to determine which of the two color coding mechanisms (trichromatic or opponent-process), better explains the color of the afterimages. We used six differently colored rectangles: red, green, blue, cyan, magenta, yellow.

We start our analysis by loading the experiment and stimuli data. The experiment data include subject index, reaction time, response in RGB format, stimuli name (e.g blue) and stimuli values in RGB and HSV. The stimuli data set includes only the information about stimuli (names, RGB and HSV values).

\begin{CodeChunk}
    \begin{CodeInput}
        R> library(bayes4psy)
        R> library(dplyr)
        R> library(ggplot2)
    
        R> data_all <- read.table("./data/after_images.csv",
                                  sep="\t", header=TRUE)
                                
        R> stimuli <- read.table("./data/after_images_stimuli.csv",
                                  sep="\t", header=TRUE)
    \end{CodeInput}
\end{CodeChunk}

We can then fit the Bayesian color model for red color stimuli and inspect the fit.

\begin{CodeChunk}
    \begin{CodeInput}
        R> data_red <- data_all %>% filter(stimuli == "red")
        R> data_red <- data.frame(r=data_red$r,
                                  g=data_red$g,
                                  b=data_red$b)
                                  
        R> fit_red <- b_color(colors=data_red)
        
        R> plot_trace(fit_red)
        R> print(fit_red)

        R> plot_fit_hsv(fit_red)       
    \end{CodeInput}
\end{CodeChunk}

\begin{figure}[ht]
    \centering
    \includegraphics[width=0.3\textwidth]{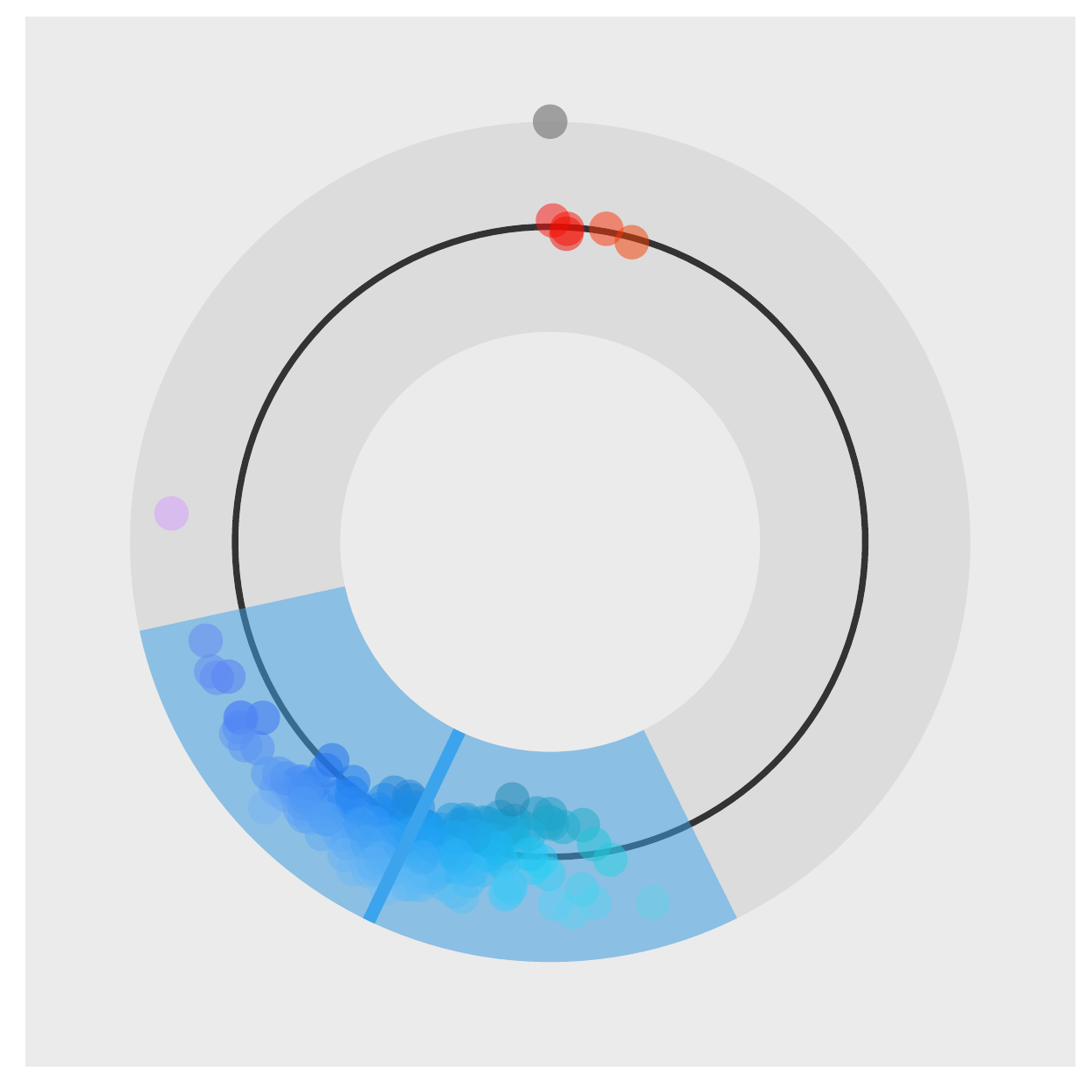}
    \caption{{\bf The special \code{plot\_fit\_hsv} function developed for the color model.} Input data points are visualized with circles, mean of the fit is visualized with a solid line and the 95\% HDI of the underlying distribution is visualized as a colored band.}
    \label{fig:ai_plot_fit_hsv}
\end{figure}

We repeat the same process five more times for the remaining five colors of stimuli. We start the analysis by loading data about the colors predicted by the trichromatic or the opponent-process theory.

\begin{CodeChunk}
    \begin{CodeInput}
        R> trichromatic <-
            read.table("./data/after_images_trichromatic.csv",
                        sep="\t", header=TRUE)
                        
        R> opponent_process <-
            read.table("./data/after_images_opponent_process.csv",
                        sep="\t", header=TRUE) 
    \end{CodeInput}
\end{CodeChunk}

We can then use the \code{plot\_distributions\_hsv} function of the Bayesian color model to produce for each stimuli a visualization of the accuracy of both color coding mechanisms predictions. Each graph visualizes the fitted distribution, displayed stimuli and responses predicted by the trichromatic and opponent-process coding. This additional information can be added to the visualization via annotation points and lines. Below is an example for the red stimulus, visualizations for other five stimuli are practically the same.

\begin{CodeChunk}
    \begin{CodeInput}
        R> stimulus <- "red"
        R> lines <- list()
        R> lines[[1]] <-
            c(trichromatic[trichromatic$stimuli == stimulus, ]$h,
              trichromatic[trichromatic$stimuli == stimulus, ]$s,
              trichromatic[trichromatic$stimuli == stimulus, ]$v)
        R> lines[[2]] <-
            c(opponent_process[opponent_process$stimuli == stimulus, ]$h,
              opponent_process[opponent_process$stimuli == stimulus, ]$s,
              opponent_process[opponent_process$stimuli == stimulus, ]$v)
        
        R> points <- list()
        R> points[[1]] <-
            c(stimuli[stimuli$stimuli == stimulus, ]$h_s,
              stimuli[stimuli$stimuli == stimulus, ]$s_s,
              stimuli[stimuli$stimuli == stimulus, ]$v_s)
        
        R> plot_red <- plot_distributions_hsv(fit_red, points=points,
                                              lines=lines, hsv=TRUE)
        R> plot_red <- plot_red + ggtitle("Red") +
                        theme(plot.title = element_text(hjust = 0.5))
    \end{CodeInput}
\end{CodeChunk}

We use the \pkg{cowplot} library to combine the plots.

\begin{CodeChunk}
    \begin{CodeInput}
        R> cowplot::plot_grid(plot_red, plot_green, plot_blue,
                             plot_yellow, plot_cyan, plot_magenta,
                             ncol=3, nrow=2, scale=0.9)
    \end{CodeInput}
\end{CodeChunk}

\begin{figure}[ht]
    \centering
    \includegraphics[width=\textwidth]{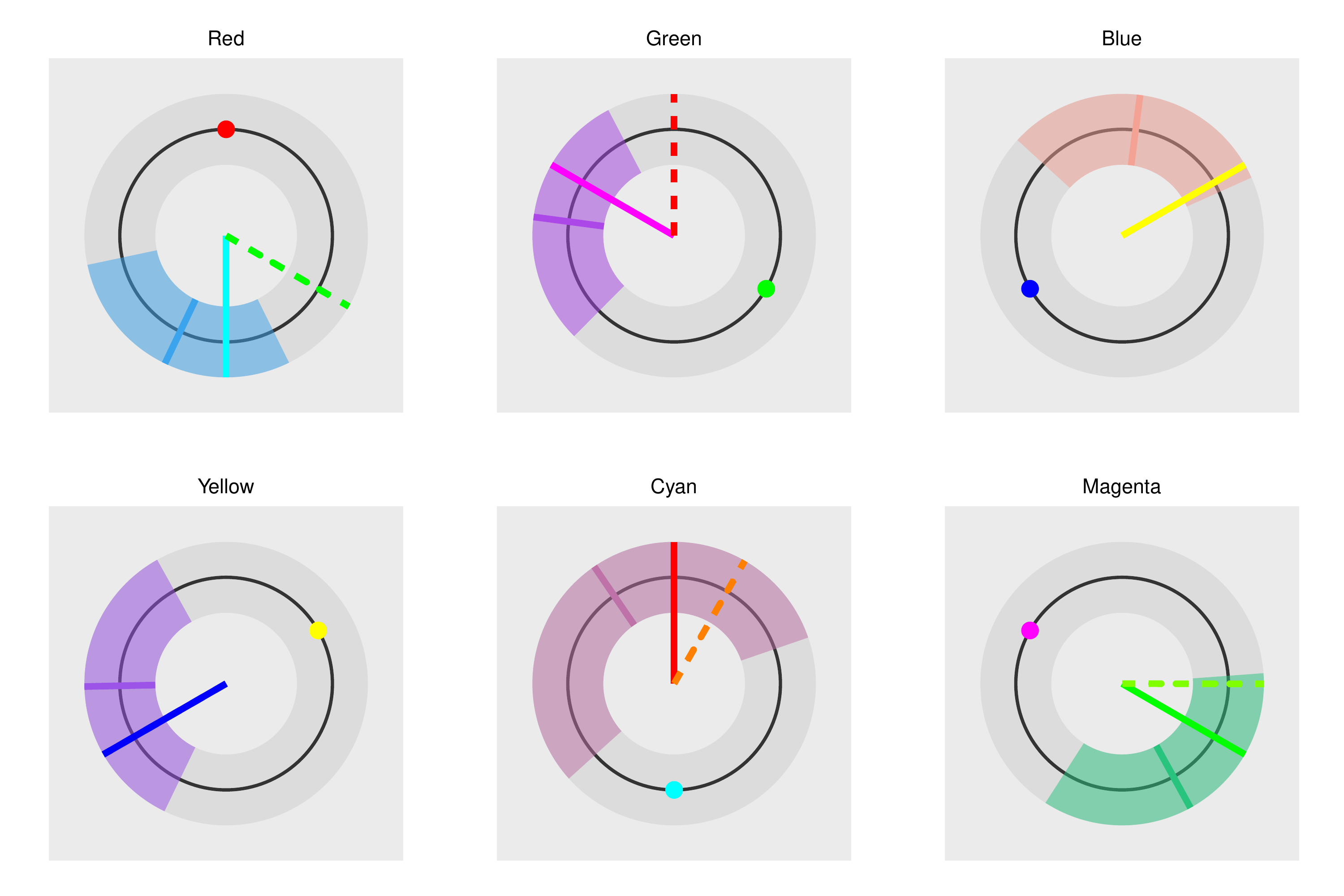}
    \caption{{\bf A comparison of thrichromatic and oponent-process color coding prediction.} The long solid line visualizes the trichromatic color coding prediction while the dashed line visualizes the opponent-process color coding. Short solid line represents the mean hue of the fit and the the colored band the 95\% HDI of the distribution underlying the fit. The small colored circle visualizes the color of the presented stimuli. In the case of blue and yellow stimuli the dashed line is not visible because both color codings predict the same outcome. The prediction based on the thrichromatic color coding seems more accurate as its prediction is always inside the 95\% of the most probable subject's responses and is always closer to the mean predicted hue than the opponent-process prediction. The opponent-process prediction is outside of the 95\% of the most probable subject's responses in cases of red and green stimuli.}
    \label{fig:plot_distributions_hsv}
\end{figure}

\section{Conclusion}
\label{sec:summary}

The \pkg{bayes4psy} package helps psychology students and researchers with little or no experience in Bayesian statistics and probabilistic programming to do modern Bayesian analysis in \proglang{R}. The package includes several Bayesian models that cover a wide range of tasks that arise in psychological experiments. Users can perform a Bayesian t-test or Bayesian bootstrap and can analyze reaction times, success rates, colors or sequential tasks. The package covers all parts of Bayesian data analysis, from fitting and diagnosing fitted models to model visualization and comparison.

We plan to upgrade the package with additional tools that will bring Bayesian statistics even closer to non-technical researchers. For example, we will implement probability distribution elicitation tools, which will ease the extraction of prior knowledge from domain experts and the prior construction process \citep{Morris2014Eliciting}. Over the last couple of years neuroimaging techniques (e.g. fMRI and EEG) have become very popular for tracking brain activity during psychological experiments. The implementation of Bayesian models for analyzing such data is also one of our future goals.

\section*{Computational details}

The results in this paper were obtained using \proglang{R}~3.5.3. \proglang{R} itself and all packages used are available from the Comprehensive \proglang{R} Archive Network (CRAN) at \url{https://CRAN.R-project.org/}.

The source code of the \pkg{bayes4psy} package can be found at \url{https://github.com/bstatcomp/bayes4psy} and the illustrative examples from Section \ref{sec:illustrations} can be found at \url{https://github.com/bstatcomp/bayes4psy_tools}. The \pkg{bayes4psy} package is currently in the final stages of CRAN publication.

\section*{Acknowledgments}

The research behind this manuscript was partially funded by the Slovenian Research Agency (ARRS) through grants L1-7542 (Advancement of computationally intensive methods for efficient modern general-purpose statistical analysis and inference), P3-0338 (Physiological mechanisms of neurological disorders and diseases), J3-9264 (Decomposing cognition: Working memory mechanism and representations), P5-0410 (Digitalization as driving force for sustainability of individuals, organizations, and society) and P5-0110 (Psychological and neuroscientific aspects of cognition).


\bibliography{Mendeley}

\newpage

\end{document}